\documentclass[useAMS,usenatbib]{mn2e}
\usepackage{epsfig}
\usepackage{float}
\usepackage[T1]{fontenc}
\usepackage{graphicx}
\usepackage{epstopdf}

\usepackage{natbib}

\usepackage{color}

\usepackage{aas_macros}

\usepackage{array}
\newcolumntype{L}[1]{>{\raggedright\let\newline\\
\arraybackslash\hspace{0pt}}m{#1}}
\newcolumntype{C}[1]{>{\centering\let\newline\\
\arraybackslash\hspace{0pt}}m{#1}}
\newcolumntype{R}[1]{>{\raggedleft\let\newline\\
\arraybackslash\hspace{0pt}}m{#1}}

\def\sigs{\mbox{$\sigma_\star$}}
\def\sige{\mbox{$\sigma_{\rm e}$}}

\def\Re{\mbox{$R_{\rm e}$}}

\def\Msun{\mbox{$M_\odot$}}
\def\Mtot{\mbox{$M_{\rm tot}$}}

\def\ML{\mbox{$M/L$}}
\def\dimf{\mbox{$\delta_{\rm IMF}$}}

\def\Mdyn{\mbox{$M_{\rm dyn}$}}

\def\mst{\mbox{$M_{\star}$}}

\def\Mvir{\mbox{$M_{\rm vir}$}}

\def\fdm{\mbox{$f_{\rm DM}$}}

\def\lsim{\mathrel{\rlap{\lower3.5pt\hbox{\hskip0.5pt$\sim$}}
    \raise0.5pt\hbox{$<$}}}                
\def\gsim{~\rlap{$>$}{\lower 1.0ex\hbox{$\sim$}}}

\def\sigAp{\mbox{$\sigma_{\rm Ap}$}}
\def\TtoSM{\mbox{$M_{\rm dyn}/\mst$}}

\def\SN{\mbox{$S/N$}}

\def\Mauto{\mbox{{\tt MAG\_AUTO}}}

\def\MErrautor{\mbox{{\tt MAGERR\_AUTO\_r}}}

\def\Fig{\mbox{Fig.~}}
\def\Figs{\mbox{Figs.~}}

\def\Sec{\mbox{Section~}}

\def\rband{\mbox{$r$-band}}
\def\gband{\mbox{$g$-band}}
\def\iband{\mbox{$i$-band}}

\def\Ksband{\mbox{$Ks$-band}}

\def\Remaj{\mbox{$R_{\rm e, maj}$}}

\def\rhostar{\mbox{$\langle \rho_{\rm \star} \rangle$}}

\def\Zsun{\mbox{$Z_{\rm \odot}$}}

\def\twodphot{\mbox{\textsc{2dphot}}}

\def\sqd{\mbox{~sq.~deg.}}

\title[Dynamics in KiDS]{The last 6 Gyr of dark matter assembly in massive galaxies from the Kilo Degree Survey}

\author[Tortora C. et al.]{\noindent
C.~Tortora$^{1}$\thanks{E-mail: ctortora@astro.rug.nl},
N.R.~Napolitano$^{2}$, N.~Roy$^{2,3}$, M.~Radovich$^{4}$,
F.~Getman$^{2}$, \and L.V.E.~Koopmans$^{1}$, G. A.~Verdoes
Kleijn$^{1}$, K.~H.~Kuijken$^{5}$
\\~\\
$^1$ Kapteyn Astronomical Institute, University of Groningen, P.O.
Box 800, 9700 AV Groningen, the Netherlands \\
$^2$ INAF -- Osservatorio Astronomico di
Capodimonte, Salita Moiariello, 16, 80131 - Napoli, Italy\\
$^3$ Dipartimento di Scienze Fisiche, Universit\`{a} di Napoli
Federico II, Compl. Univ. Monte S. Angelo, 80126 - Napoli, Italy\\
$^4$ INAF -- Osservatorio Astronomico di Padova, Vicolo Osservatorio 5, 35122 - Padova, Italy\\
$^5$ Leiden Observatory, Leiden University, P.O. Box 9513, 2300 RA
Leiden, the Netherlands\\}

\begin{document}
\date{Accepted  Received }
\pagerange{\pageref{firstpage}--\pageref{lastpage}} \pubyear{xxxx}
\maketitle

\label{firstpage}
\begin{abstract}
We study the dark matter (DM) assembly in the central regions of
massive early-type galaxies up to $z\sim 0.65$. We use a sample of
$\sim 3800$ massive ($\log \mst/\Msun > 11.2$) galaxies with
photometry and structural parameters from 156\sqd\ of the Kilo
Degree Survey, and spectroscopic redshifts and velocity
dispersions from SDSS. We obtain central total-to-stellar mass
ratios, \TtoSM, and DM fractions, by determining dynamical masses,
\Mdyn, from Jeans modelling of SDSS aperture velocity dispersions
and stellar masses, \mst, from KiDS galaxy colours. We first show
how the central DM fraction correlates with structural parameters,
mass and density proxies, and demonstrate that most of the local
correlations are still observed up to $z \sim 0.65$; at fixed
\mst, local galaxies have larger DM fraction, on average, than
their counterparts at larger redshift. We also interpret these
trends with a non universal Initial Mass Function (IMF), finding a
strong evolution with redshift, which contrast independent
observations and is at odds with the effect of galaxy mergers. For
a fixed IMF, the galaxy assembly can be explained, realistically,
by mass and size accretion, which can be physically achieved by a
series of minor mergers. We reproduce both the \Re--\mst\ and
\TtoSM--\mst\ evolution with stellar and dark mass changing at a
different rate. This result suggests that the main progenitor
galaxy is merging with less massive systems, characterized by a
smaller \TtoSM, consistently with results from halo abundance
matching.
\end{abstract}


\section{Introduction}\label{sec:intro}

Dark matter (DM) dominates the mass density of galaxies and
clusters of galaxies. Its budget amounts to $\sim 85$ per cent of
the total mass density of the universe (e.g.,
\citealt{SDSS_DR1,SDSS_DR6,SDSS_DR7_Abazajian}) and its imprint is
found at cosmological scales along all cosmic history (e.g.,
\citealt{Komatsu+11_WMAP7}). The strongest constraints on the
shapes and properties of DM haloes come from numerical simulations
of (DM only) structure formation within the consensus cosmology
framework, i.e. the $\Lambda$CDM model (\citealt{NFW96}, hereafter
NFW; \citealt{Bullock+01}; \citealt{Maccio+08}). However, more
realistic models, which try to account for the effect of baryons
on the DM distribution (e.g., \citealt{Blumenthal+84_nature};
\citealt{Gnedin+04}; \citealt{Wu+14}) seem more compatible with
observations (e.g. \citealt{Gnedin+07}; \citealt{NRT10}) and make
more realistic predictions on the expected DM fractions in the
central galaxy regions (\citealt{RS09}; \citealt{Hilz+13};
\citealt{Wu+14})

Early-type galaxies (ETGs, ellipticals and lenticulars) contain
most of the cosmic stellar mass of the universe, and represent the
final stage of galaxy evolution. They hold the fossil record of
the stellar and DM assembly through time, and, being the most
luminous and massive galaxies, can be studied in details out to
large redshifts. In particular, since ETGs are thought to be the
product of the transformation of late-type galaxies' (LTGs), e.g.
through the effect of merging and other feedback mechanisms, they
are crucial to understand the processes that shape galaxies across
time.

In this context it is important to trace the assembly of both the
luminous and the dark components of these systems. E.g., the total
stellar-to-dark mass ratio of ETGs depends strongly on the galaxy
mass, and seems to be connected to the overall star formation
efficiency (\citealt{Benson+00}; \citealt{MH02};
\citealt{Napolitano+05}; \citealt{Mandelbaum+06};
\citealt{vdB+07}; \citealt{CW09}; \citealt{Moster+10};
\citealt{Alabi+16}).

But DM rules also the central galaxy regions
(\citealt{Gerhard+01}; \citealt{Padmanabhan+04};
\citealt{Cappellari+06}; \citealt{ThomasJ+07};
\citealt{Cardone+09}; \citealt{ThomasJ+09}; \citealt{HB09_FP};
\citealt{Tortora+09}; \citealt{Auger+10_SLACSX}; \citealt{CT10};
\citealt{ThomasJ+11}; \citealt{Cardone+11SIM};
\citealt{SPIDER-VI}; \citealt{Tortora+14_DMevol};
\citealt{Shu+15_SLACSXII}; \citealt{Nigoche-Netro+16}), in a way
substantially consistent with the concordance $\Lambda$CDM
scenario (\citealt{Tortora+09}; \citealt{NRT10};
\citealt{SPIDER-VI}). Different works have shown that the central
DM fraction (typically within one effective radius, \Re\
hereafter) is higher in larger and more massive galaxies (e.g.
\citealt{HB09_FP}; \citealt{Tortora+09}; \citealt{RS09};
\citealt{Auger+10_SLACSX}; \citealt{NRT10}; \citealt{ThomasJ+11};
\citealt{SPIDER-VI}), even though there is no universal consensus
about such a trend as also evidences of an anti-correlation with
mass have been presented (e.g., \citealt{Grillo+09, Grillo10,
Grillo_Cobat10}).

The claimed correlation with mass seems almost insensitive to the
adopted galaxy mass profile or initial mass function, IMF (e.g.,
\citealt{Cardone+09}; \citealt{CT10}; \citealt{Cardone+11SIM}),
but it can become uncertain in case a non-$\Lambda$CDM scenario,
with mass following the (non-homologous) light distribution, is
adopted (e.g., \citealt{TBB04}; \citealt{Tortora+09,SPIDER-VI}).
The scenario is even more complicated if one takes into account
the effect of a non universal IMF (\citealt{vDC10};
\citealt{Treu+10}; \citealt{ThomasJ+11};
\citealt{Conroy_vanDokkum12b}; \citealt{Cappellari+12,
Cappellari+13_ATLAS3D_XX}; \citealt{Spiniello+12};
\citealt{Wegner+12}; \citealt{Barnabe+13}; \citealt{Dutton+13};
\citealt{Ferreras+13}; \citealt{Goudfrooij_Kruijssen13};
\citealt{LaBarbera+13_SPIDERVIII_IMF}; \citealt{TRN13_SPIDER_IMF};
\citealt{Weidner+13_giant_ell}; \citealt{Goudfrooij_Kruijssen14};
\citealt{Shu+15_SLACSXII}; \citealt{McDermid+14_IMF};
\citealt{Tortora+14_DMslope,Tortora+14_MOND};
\citealt{Martin-Navarro+15_IMF_variation};
\citealt{Spiniello+15_IMF_vs_density}; \citealt{Lyubenova+16};
\citealt{TLBN16_IMF_dwarfs}; \citealt{Corsini+17};
\citealt{Li+17_IMF}; \citealt{Sonnenfeld+17_IMF};
\citealt{Tortora+17_Verlinde}). Indeed, the IMF remains the
largest source of uncertainty to quantify the stellar and DM mass
budget in the central galactic regions. In absence of direct
constraints (e.g. using gravity sensitive spectral lines, see
\citealt{Spiniello+12}; \citealt{LaBarbera+13_SPIDERVIII_IMF}),
the adoption of different ``universal'' IMF recipes causes the
stellar mass to vary by a factor of 2 or more (i.e. assuming a
\citealt{Chabrier01} or a \citealt{Salpeter55} IMF or even
super-Salpeter IMF, e.g. \citealt{Tortora+09}) hence strongly
affecting the conclusions on the central DM fraction in these
extreme cases. In case of ``non-universality'', the systematic
variation of the IMF with mass (or velocity dispersion), from a
bottom-lighter (i.e., 'lower-mass') IMF for low mass systems to a
bottom-heavier (i.e., 'higher-mass') IMF in massive galaxies could
dilute (and even cancel) the ``apparent'' DM fraction trend with
mass (e.g., \citealt{ThomasJ+11}; \citealt{TRN13_SPIDER_IMF};
Spiniello's thesis, Chapter 2). However, once again, the scenario
is far to be fully constrained as for the most massive galaxies
some contrasting results point to unexpected low stellar
mass-to-light ratios (\ML s) and bottom-light IMFs have been found
(\citealt{Smith+15_SINFONI}).

One way to interpret all these evidences in the context of the
galaxy evolution is to check the persistence of these correlations
at higher-redshift and find the epochs where these effects start
to emerge. This implies a test on the assembly of both the dark
and the stellar matter in galaxies, at epochs where a) both are in
an early stage of their evolution and b) the freedom on some
parameters (e.g. age, metallicity of stars, concentration of the
DM haloes, etc.) is minimal.

In order to cover the full parameter space, including the
look-back time, we need large galaxy samples. So far, most of the
DM studies were limited to low-redshift samples, and only recently
systematic analysis of high redshift samples have been started. In
some cases the datasets are restricted to small samples and small
redshift windows to evaluate the dependence of the galaxy DM
content on redshift, as in the case of gravitational lenses
(\citealt{Auger+09_SLACSIX, Auger+10_SLACSX},
\citealt{Tortora+10lensing}; \citealt{Sonnenfeld+13_SL2S_IV}). The
first studies have given contrasting results (\citealt{Faure+11};
\citealt{Ruff+11}). The reason of the tensions among these latter
studies probably resides in the paucity of the galaxy samples and
differences in the model choices.

The first systematic studies of the evolution of the central DM
fraction with redshift has been recently performed by
\cite{Beifiori+14} and \cite{Tortora+14_DMevol}, which provided
evidences that high$-z$ ETGs are less DM dominated than their
local counterparts.

However, this line of investigations has just started and further
independent analyses are needed, not only to constrain the overall
evolution of central DM, but also to assess the correlations with
structural parameters, mass and stellar density, and evaluate how
these change as a function of redshift, within or not the
non-universal IMF scenario. To make even a step forward into the
previous analysis, we have applied the Jeans method discussed in
\cite{Tortora+14_DMevol} to a state-of-the-art sample of galaxies
covering a broad redshift range for which high quality imaging and
internal kinematics were available, both necessary to characterize
the stellar and total mass for these systems. We have gathered a
sample of massive galaxies with high-quality imaging, measured
structural parameters and stellar masses from the Kilo Degree
Survey (KiDS). KiDS is one of the public surveys carried out with
the VST telescope, which is characterized by the excellent image
quality, thanks to the very good seeing ($0.65$ arcsec, on
average, in the \rband) and a high depth in the $r-$band ($\sim25$
mag limiting magnitude). The KiDS fields in the Northern galaxy
cap partially overlap with SDSS--DR7 data sample and with
BOSS@SDSS, which both provided the spectroscopic redshifts and
central velocity dispersions for our galaxy sample. Jeans
modelling was used to determine dynamical masses and
total-to-stellar mass ratios, to be correlated with galaxy
parameters and redshift. Our results are also compared with those
from a) low-redshift ($0.05 < z < 0.095$) ETGs from the SPIDER
(Spheroid's Panchromatic Investigation in Different Environmental
Regimes) project (\citealt{SPIDER-I}; \citealt{SPIDER-VI}), b) a
spectroscopically selected sample of ETGs covering the range of
redshifts $z\sim 0.4-0.8$ from the ESO Distant Clusters Survey
(EDisCS; \citealt{Saglia+10}; \citealt{Tortora+14_DMevol}), and c)
other results from literature observations and simulations.

The paper is organized as follows. Data samples and the analysis
performed are presented in \Sec\ref{sec_data_analysis}. The
correlation with structural parameters, mass probe and stellar
density are discussed in \Sec\ref{sec:DM_correlations}.
\Sec\ref{sec:DM_evolution} is devoted to the systematic analysis
of central DM and IMF evolution with redshift, systematics and the
interpretation within the merging scenario. A summary of the
results, conclusions and future prospects are discussed in
\Sec\ref{sec:conclusions}. We adopt a cosmological model with
$(\Omega_{m},\Omega_{\Lambda},h)=(0.3,0.7,0.75)$, where $h =
H_{0}/100 \, \textrm{km} \, \textrm{s}^{-1} \, \textrm{Mpc}^{-1}$
(\citealt{Komatsu+11_WMAP7}).

\section{Analysis}\label{sec_data_analysis}

\subsection{KiDS and SDSS datasamples}

The galaxy sample presented in this work is selected from the data
included in the first, second and third data releases of KiDS
presented in \cite{deJong+15_KiDS_paperI} and
\cite{deJong+17_KiDS_DR3}. The total dataset includes 156 KiDS
pointings with the measured structural parameters presented in Roy
et al. (2017, in preparation). We have identified about 22 million
sources, including $\sim$7 million which have been classified as
high quality extended sources. We select those systems with the
highest \SN\ in the \rband\ images, $\SN_r \equiv$ 1/\MErrautor $>
50$, with reliable structural parameters measured. This dataset
includes aperture and total photometry, photometric redshifts and
structural parameters.

To record spectral information, such as spectroscopic redshifts
and velocity dispersions, this data-sample is cross-matched with
two different SDSS samples, collecting a sample of galaxies with
redshifts in the range $0 < z < 0.7$:
\begin{itemize}
\item {\it MPA-JHU-DR7}. For the lowest redshifts ($z < 0.2$) we base our analysis
on the spectroscopic data from the seventh Data Release of the
SDSS (DR7; \citealt{SDSS_DR7_Abazajian}). In particular, we select
these systems, getting redshifts and velocity dispersions from the
MPA-JHU-DR7 catalog\footnote{The data catalogs are available from
{\tt http://wwwmpa.mpa-garching.mpg.de/SDSS/DR7/raw\_data.html}.},
which consists of $\sim 928000$ galaxies of any type with
redshifts $z \lsim 0.7$. The cut in mass which we will perform
later will remove almost all the late-type contaminants. Spectra
are measured within fibers of diameter 3 arcsec.
\item {\it BOSS-DR10}. Data at redshift $z \geq 0.2$ are taken from the
SDSS-III/BOSS Data Release Ten\footnote{The data catalogs are
available from {\tt
http://www.sdss3.org/dr10/spectro/galaxy\_portsmouth.php}.} (DR10,
\citealt{Ahn+14_SDSS_DR10}). Selection criteria are designed to
identify a sample of luminous and massive galaxies with an
approximately uniform distribution of stellar masses following the
Luminous Red Galaxy (LRG; \citealt{Eisenstein+11_SDSSIII}) models
of \cite{Maraston+09_LRG}. The galaxy sample is composed of two
populations: the higher-redshift Constant Mass Sample (CMASS; $0.4
< z < 0.7$) and the Low-Redshift Sample (LOWZ; $0.2 < z < 0.4$).
The total sample, which consists of 934000 spectra and velocity
dispersions across the full SDSS area, starts to be incomplete at
redshift $z \gsim 0.6$ and masses $\log \mst/\Msun \gsim 11.3$.
The fiber diameter is of 2 arcsec. Velocity dispersions are
determined in \cite{Thomas+13_BOSS}, using Penalized PiXel Fitting
(pPXF, \citealt{Cappellari_Emsellem14}) and GANDALF
(\citealt{Sarzi+06_SAURONV}) on the BOSS spectra. These values are
quite robust being, on average, quite similar to the measurements
from independent literature (see \citealt{Thomas+13_BOSS} for
further details).
\end{itemize}

The final sample consists of 4118 MPA-JHU-DR7 galaxies and 5603
BOSS-DR10 galaxies, for a total of 9721 systems with structural
parameters, spectroscopic redshifts and velocity dispersions. We
limit to a mass-completed sample of galaxies with $\log \mst/\Msun
> 11.2$, consisting of a total of 3778 galaxies with redshift $0 <
z < 0.7$.

In the following subsections we will provide more details about
the products of the analysis of the KiDS dataset and the dynamical
procedure. In particular, in \Sec\ref{subsec:struc} we will
describe how the structural parameters are determined. In
\Sec\ref{subsec:st_masses} we will provide details about the
derivation of the stellar masses and the dynamical Jeans modelling
is discussed in \Sec\ref{subsec:dynamics}. In
\Sec\ref{subsec:DM_and_rest-frame} we will define the
total-to-stellar mass ratio and DM fraction. Finally, in
\Sec\ref{subsec:progenitor_bias} we discuss how progenitor bias is
taken into account.

\subsection{Structural parameters}\label{subsec:struc}

Galaxy structural parameters have been derived via accurate 2D
surface photometry of the highest \SN\ sample
(\citealt{LaBarbera_08_2DPHOT, SPIDER-I}; Roy et al. 2017, in
preparation). Surface photometry is performed using the \twodphot\
environment, an automatic computer code designed to obtain both
integrated and surface photometry of galaxies in wide-field
images. The software first produces a local PSF model from a
series of identified {\it sure stars}. For each galaxy, this is
done by fitting the four closest stars to that galaxy with a sum
of three two-dimensional Moffat functions. Then galaxy snapshots
are fitted with PSF-convolved S\'ersic models having elliptical
isophotes plus a local background value (see
\citealt{LaBarbera_08_2DPHOT} for further details). The fit
provides the following parameters for the four wavebands: surface
brightness at \Re, $\mu_{\rm e}$, circularized effective radius,
\Re, S\'ersic index, $n$, total magnitude, $m_{S}$, axis ratio,
$q$, and position angle. As it is common use in the literature, in
the paper we use the circularized effective radius, \Re, defined
as $\Re = \sqrt{q} \Remaj$, where \Remaj\ is the major-axis
effective radius. For further details about the catalog extraction
and data analysis see Roy et al., in preparation.

\subsection{Stellar mass determination}\label{subsec:st_masses}

To determine stellar masses, \mst , we have used the software
\textsc{le phare} (\citealt{Arnouts+99}; \citealt{Ilbert+06}),
which performs a $\chi^{2}$ fitting method between the stellar
population synthesis (SPS) theoretical models and data. Single
burst models from \cite{BC03}, with different metallicities ($0.2
\leq Z/\Zsun \leq 2.5$) and ages ($3 \leq age \leq \rm age_{\rm
max}$ Gyr), and a \cite{Chabrier01} IMF is used. The
\cite{Salpeter55} gives masses larger of a factor $\sim 1.8$
(\citealt{Tortora+09}; \citealt{SPIDER-V}). The maximum age, $\rm
age_{\rm max}$, is set by the age of the Universe at the redshift
of the galaxy, with a maximum value at $z=0$ of $13\, \rm Gyr$. To
minimize the probability of underestimating the stellar mass by
obtaining too low an age, following \cite{Maraston+13_BOSS} we
have applied age cutoffs to the model templates, allowing for a
minimum age of 3 Gyr. Models are redshifted using the SDSS
spectroscopic redshifts. We adopt the observed KiDS $ugri$
photometry (and related $1\, \sigma$ uncertainties) within a $6''$
aperture of diameter, corrected for Galactic extinction using the
map in \cite{Schlafly_Finkbeiner11}. Total magnitudes derived from
the S\'ersic fitting, $m_{S}$, are used to correct the outcomes of
\textsc{le phare} for missing flux. The single burst assumption,
as well as the older stellar populations and metal-richer models
are suitable to describe the red and massive galaxies we are
interested in (\citealt{Gallazzi+05}; \citealt{Thomas+05};
\citealt{Tortora+09}). Among \textsc{le phare} outputs, we will
adopt best-fitted masses in this paper.

\subsection{Dynamical modelling}\label{subsec:dynamics}

Following the analysis in \cite{Tortora+09} and \cite{SPIDER-VI}
we model the SDSS aperture velocity dispersion of individual
galaxies using the spherical isotropic Jeans equations to
estimate the (total) dynamical mass \Mdyn\ (which, we will also
refer to as \Mtot) within $r=$~1~\Re. In the Jeans equations, the
stellar density and the total mass distribution need to be
specified. Thus, the stellar density is provided by the 2D S\'ersic
fit of the KiDS $r$-band galaxy images, and the total (DM + stars) mass is
assumed to have the form of a Singular Isothermal Sphere (SIS),
from which $M(r)\propto \sigma_{\rm SIS}^{2}  r$ (corresponding to
a 3D mass density slope $\gamma = 2$), where $\sigma_{\rm  SIS}$
is the model (3D) velocity dispersion.

The total mass density profile in the centre of ETGs flattens with
galaxy mass (\citealt{Remus+13}; \citealt{Dutton_Treu14};
\citealt{Tortora+14_DMslope}; \citealt{Poci+17_slope}): low-mass
ETGs have steep mass density distributions consistent with those
of stars (i.e. consistently with a constant-\ML\ profiles), while
shallower isothermal profiles has been found to provide a robust
description of the mass distribution in massive ETGs (e.g.,
\citealt{Kochanek91}; \citealt{Bolton+06_SLACSI};
\citealt{Koopmans+06_SLACSIII}; \citealt{Gavazzi+07_SLACSIV};
\citealt{Bolton+08_SLACSV}; \citealt{Auger+09_SLACSIX};
\citealt{Auger+10_SLACSX}; \citealt{Chae+14}; \citealt{Oguri+14}).
This ``conspiracy'' (\citealt{Rusin+03, TK04,
Koopmans+06_SLACSIII, Gavazzi+07_SLACSIV}; \citealt{Tortora+09};
\citealt{Auger+10_SLACSX}; \citealt{Tortora+14_DMslope}) seems to
be motivated also by theoretical arguments: an overall isothermal
profile can be explained by a smaller amount of dissipation during
the formation of such high-mass galaxies, if compared to
lower-mass systems, where higher level of dissipation leads to a
more prominent contribution from newly formed stars to the total
mass density in the center, steepening their total density slope.
(\citealt{Koopmans+06_SLACSIII}; \citealt{Remus+13};
\citealt{Tortora+14_DMslope}; \citealt{Remus+17}). For further
details on the systematics introduced by the particular mass
density profile choice, one can refer to \cite{Tortora+09} and
\cite{SPIDER-VI} (see also
\citealt{Cardone+09,CT10,Cardone+11SIM}).

We will discuss the impact of a non-isothermal mass density
profile and orbital anisotropy on our inferences in
\Sec\ref{subsec:systematics}.

\subsection{Dark matter content and rest-frame quantities}\label{subsec:DM_and_rest-frame}

We characterize the mass content of an ETG by computing the
de-projected total-to-stellar mass ratio $\Mdyn(r)/\mst(r)$, i.e.
the ratio between dynamical and stellar mass in a sphere of radius
$r$ and refer to the value assumed by this quantity at the
effective radius \Re, i.e. $\Mdyn(\Re)/\mst(\Re)$, as the
``central'' total-to-stellar mass ratio. As the total dynamical
mass includes both stars and DM, we will also use a related
quantity, which makes explicit the DM fraction within \Re, defined
as $\fdm(\Re) = 1 - \mst(\Re) / \Mdyn(\Re)$. When not stated
explicitly, \mst\ is referred to the SPS value assuming a Chabrier
IMF, discussed in \Sec\ref{subsec:st_masses}. Note that usually
dynamical analysis formalisms include de-projected masses (e.g.
see \citealt{Tortora+09}), while projected masses are typically
present in strong lensing equations (e.g.
\citealt{Auger+10_SLACSX}). The projected \TtoSM\ or \fdm\ are
always larger than their de-projected versions within the same
radius, because of the contribution of the outer parts of the halo
along the line-of-sight\footnote{The projected stellar mass within
\Re\ is $0.5 \times \mst$, while the de-projected stellar mass
within the same radius is about $0.416 \times \mst$ (calculated
using a S\'ersic profile with $n=4$). Instead, for a SIS, the
projected mass is $\pi/2$ times the spherical mass, and this value
is constant with radius. Therefore, the projected total-to-stellar
mass ratio is $\sim 1.3$ times (i.e. $\sim 0.12$ dex) larger than
the equivalent de-projected quantity.}. In the following, we will
discuss the DM quantities described above as a function of
structural parameters, masses, velocity dispersion, stellar
density and, mainly, as a function of the redshift.

In the spherical Jeans equation, for the stellar density we have
adopted the observed \rband\ structural parameters, while
$\Mdyn(\Re)/\mst(\Re)$ are computed using rest-framed $n$ and \Re.
Indeed, the effective radii should be referred to a fixed
rest-frame wavelength to account for the effect of color
gradients, which make ETG optical \Re\ larger in bluer than in
redder bands, on average. If this effect is not taken into
account, then \Re\ are systematically larger at higher redshift
(\citealt{Sparks_Jorgensen93}; \citealt{HB09_curv};
\citealt{LaBarbera_deCarvalho09}; \citealt{Roche+10};
\citealt{Beifiori+14}; \citealt{Tortora+14_DMevol};
\citealt{Vulcani+14}). Similar considerations hold for the
S\'ersic indices. In particular, using a sample of galaxies with
$z < 0.3$, \cite{Vulcani+14} estimate an increase from $g$ to $u$
and from $r$ to $g$-band of $\lsim 15$ per cent, and similar
results are found in \cite{LaBarbera_deCarvalho09} following the
method in \cite{Sparks_Jorgensen93}. We have determined the
rest-frame structural parameters ($X=\Re$ and $n$) by
interpolating the observed X parameters in the three KiDS
wavebands $g$, $r$ and $i$. We have performed a linear fit
\begin{equation}
\log X = a + b \log \lambda,
\label{eq:rest_X}
\end{equation}
to the data points $(\lambda_{\rm l}, X_{\rm l})$, with $l=g,r,i$,
where $\lambda_{g,r,i} = \{ 4735, 6287, 7551\}$ \AA\ are the mean
wavelengths of our filters. Then, we have assumed the $g$-band
structural parameters at $z=0$, $X(\lambda_{g})$ in Eq.
\ref{eq:rest_X}, and calculated the rest-frame $g$-band structural
parameters at $z>0$ as $X((1+z)\lambda_{g})$. The average shifts
with respect to the \gband\ quantities are $-5$ per cent for \Re\
and $7$ per cent for $n$, with scatter of $50$ and $40$ per cent,
respectively.

\subsection{Progenitor bias}\label{subsec:progenitor_bias}

The results need to be corrected for progenitor bias, since
low-$z$ ETG samples contain galaxies that have stopped their star
formation only recently and that would not be recognized as ETGs
at higher redshifts. This is the case of systems with relatively
young ages that cannot correspond to passive objects at higher-$z$
(\citealt{vanDokkum_Franx01}; \citealt{Saglia+10};
\citealt{Valentinuzzi+10_WINGS, Valentinuzzi+10_EDisCS};
\citealt{Beifiori+14}; \citealt{Tortora+14_DMevol}).

The impact of the progenitor bias can push galaxy parameters in
different directions, as discussed in \cite{Tortora+14_DMevol}.
E.g., the correlation of \Re\ with galaxy age is still
controversial. In fact, contrasting results are found by
observational analysis, which show that, at fixed mass or velocity
dispersion, younger systems are larger
(\citealt{Shankar_Bernardi09}; \citealt{NRT10};
\citealt{Tortora+10lensing}; \citealt{Valentinuzzi+10_WINGS})
or are as sized as older galaxies (\citealt{Graves+09},
\citealt{Tortora+14_DMevol}).
The outcomes from semi-analytic
galaxy formation models are also still unclear, as there are
results showing that younger galaxies are larger
(\citealt{Khochfar_Silk06}) or also smaller (\citealt{Shankar+10})
than the oldest systems.

To correct for the progenitor bias we would need an accurate
estimate of the galaxy ages. Unfortunately our galaxy age have
been obtained from the fitting of spectral models to our KiDS
optical multi-band photometry, hence they cannot be more than a
qualitative guess. Thus, following \cite{Beifiori+14}, we remove
those galaxies whose age at redshift $z = 0.65$ -- the centre of
our highest redshift bin -- was less than 3 Gyr, which is the time
needed for a typical galaxy to become passive. This cut leaves
2595 galaxies, i.e. about $69$ per cent of the total sample. In
the rest of the paper, because of the uncertainties in our
photometric ages, we will discuss both the results without and
with this progenitor bias correction.

\section{Correlation with structural parameters and mass
probes}\label{sec:DM_correlations}

\begin{figure*}
\psfig{file=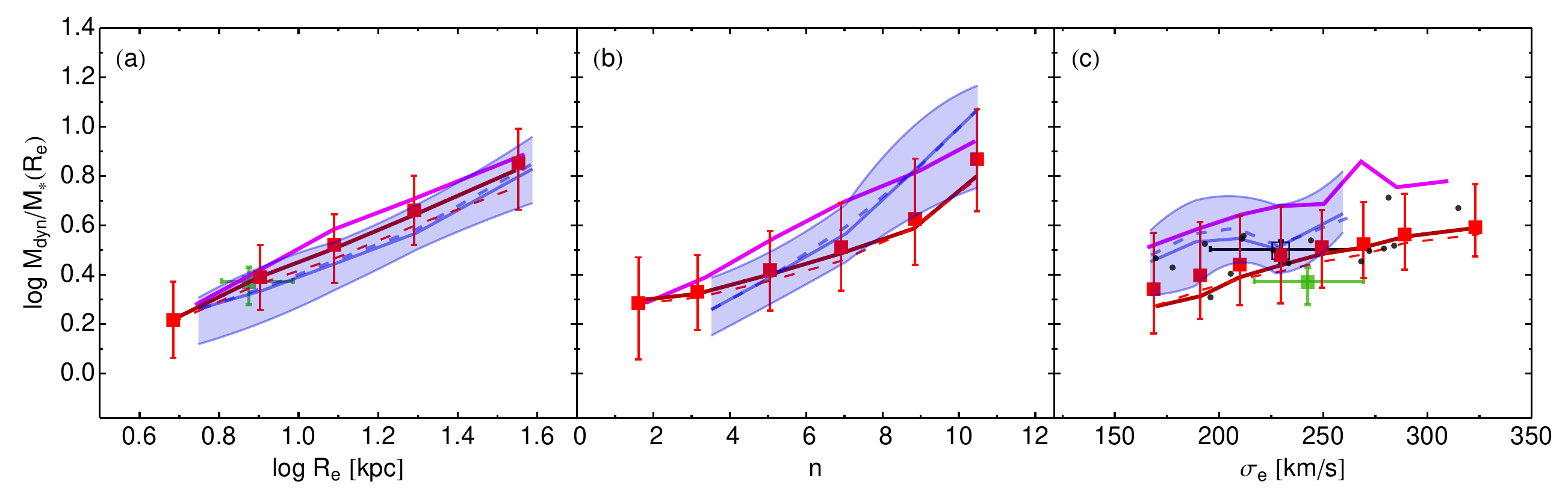, width=1.00\textwidth}
\psfig{file=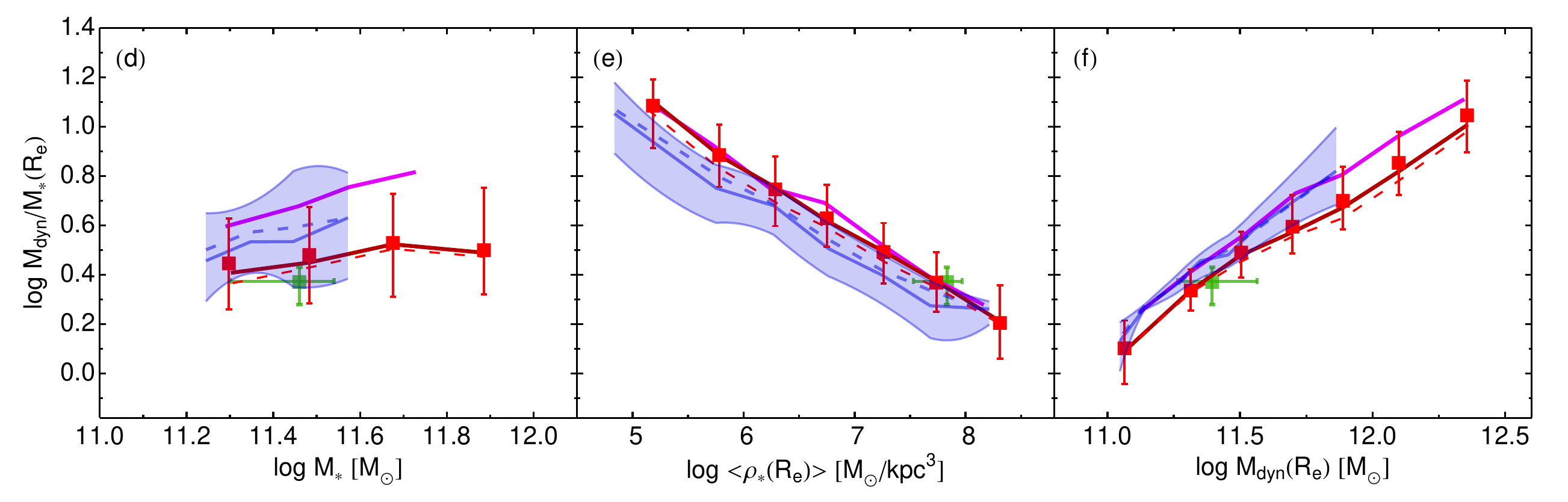, width=1.00\textwidth}
\caption{The total-to-stellar mass ratio $\Mdyn/\mst$ within
rest-frame effective radius, \Re, assuming a Chabrier IMF, as a
function of (a) rest-frame effective radius \Re, (b) rest-frame
S\'ersic index n, (c) velocity dispersion within rest-frame
effective radius $\sigma_{\rm e}$, (d) total stellar mass \mst,
(e) central average stellar density \rhostar\ and (f) dynamical
mass \Mdyn within rest-frame \Re, $\Mdyn(\Re)$. The de-projected
S\'ersic law in the rest-frame $g$-band is used to describe the
density profile of the stellar component. Red squares and error
bars are medians and 25--75th percentile trends for the whole KiDS
sample under analysis. Solid purple and dark red lines are medians
for galaxies in the redshift bins $0.1 < z \leq 0.3$ and $0.3 < z
\leq 0.7$. Dark blue line and light blue region are medians and
25--75th percentile trends for SPIDER galaxies with $\mst >
10^{11.2}\, \rm \Msun$. Dashed red (blue) lines are medians for
results corrected for progenitor bias for KiDS (SPIDER) data-sets.
Green squares and error bars are medians and 25--75th percentiles
for SLACS lenses from \citet{Auger+10_SLACSX}. In panel (c), black
dots, open square and error bars are single datapoints, median and
25--75th percentiles for the results in \citet{ThomasJ+11}, which
apply a Schwarzschild's orbit superposition technique to a sample
of 16 COMA ETGs.}\label{fig:DM}
\end{figure*}

\subsection{Dark matter fraction}

\Fig\ref{fig:DM} shows central \TtoSM\ as a function of different
galaxy parameters, i.e. effective radius, S\'ersic index, velocity
dispersion, stellar and dynamical  mass,  and central average
de-projected stellar density, \rhostar, defined as $\rhostar =
\mst(\Re)/(\frac{4}{3}\pi \Re^3)$. Since here we have fixed the
IMF to Chabrier, the \TtoSM\ trend implies a variation in the DM
content. Red symbols are for the KiDS sample, where we have
collected all the galaxies with redshift $z < 0.7$. Dashed blue
lines with light blue shaded regions are for a sample of ETGs with
redshifts $0.05 < z < 0.095$ from the SPIDER survey, assuming
$g$-band structural parameters. Error bars and the shaded regions
are the 25--75th per cent quantiles. We will also fit the
power-law relation $\TtoSM \propto X^{\alpha}$, where $X$ is one
of the galaxy parameters (\Re, n, \sige, \mst, \rhostar, \Mdyn)
and $\alpha$ is the slope of the correlation\footnote{The \sige\
is the SDSS-fibre velocity dispersion, \sigAp , corrected to an
aperture of one \Re , following \cite{Cappellari+06}.}. All the
correlations are significant at more than 99 per cent.

We find a tight and positive correlation with a slope
$\alpha=0.72$ between \TtoSM\ and \Re, which is interpreted as a
physical aperture effect, where a larger \Re\ subtends a larger
portion of a galaxy DM halo. A similar steep correlation also
holds between \TtoSM\ and S\'ersic index ($\TtoSM \propto
n^{0.62}$), which means that galaxies with steeper light profiles
have higher central DM fractions. The galaxies with the smallest
\Re\ ($\sim 5 \, \rm kpc$) and S\'ersic indices ($\sim 2$) have
the smallest DM fraction ($\sim 35$ per cent), while the largest
galaxies ($\Re \sim 35 \, \rm kpc$) with the steepest light
profiles ($n \sim 10$) present the largest DM content ($\sim 85$
per cent).

We also find that \TtoSM\ correlates with \sige\ ($\TtoSM \propto
\sige^{0.89}$). Galaxies with larger \Mdyn, i.e. with a larger
content of both stellar and dark matter, have a larger DM content
($\TtoSM \sim 10$, i.e. $90$ per cent of DM), the slope of the
correlation is $0.7$. The correlation with \mst\ is shallower,
with an average $\TtoSM \sim 3$ (i.e. $67$ per cent of DM) and
$\alpha = 0.11$.

\Fig\ref{fig:DM} also shows a sharp anti-correlation between DM
content and central average stellar density with $\alpha=-0.28$,
which has been reported for the first time in \cite{SPIDER-VI},
and now is confirmed using samples of intermediate-redshift
galaxies. Galaxies with denser stellar cores have lower DM
fractions (i.e. $\TtoSM \sim 2$ or $\fdm \sim 0.5$ at $\rho_\star
\sim \times 10^{8} \, \rm \Msun kpc^{-3}$), while \fdm\ values as
high as $\sim 0.95$ are found at the lowest densities ($\rho_\star
\sim 10^{5} \, \rm \Msun kpc^{-3}$). This trend results from the
fact that, on average, higher stellar densities correspond to
smaller effective radii, implying a lower \TtoSM. All these trends
are qualitatively consistent with those found for a sample of
massive $z\sim 0$ SPIDER galaxies (blue lines with shaded regions;
see also \citealt{SPIDER-VI}). Our results confirm most of the
previous literature at $z \sim 0$ (e.g., \citealt{Padmanabhan+04};
\citealt{Cappellari+06}; \citealt{HB09_curv};
\citealt{Tortora+09}; \citealt{NRT10}; \citealt{SPIDER-VI}), or at
intermediate redshift (\citealt{Tortora+10lensing};
\citealt{Auger+10_SLACSX}; \citealt{Tortora+14_DMevol}).

In \Fig\ref{fig:DM} we also plot the results when progenitor bias
is accounted for (dashed lines), showing that the trends are
almost unaffected. We have finally plotted the results for two
redshift bins: $0.1 < z \leq 0.3$ (purple solid line) and $0.3 < z
\leq 0.7$ (darker red line). Excepted for the correlations with
\Re\ and \rhostar, we find an evolution in the \TtoSM, with larger
DM fraction in the lower redshift bin. Note that the median
\TtoSM\ of the $z\sim 0$ SPIDER galaxies are smaller than those of
KiDS galaxies in the lower redshift bin with $0.1 < z \leq 0.3$.
This seems to contract the trend of higher \TtoSM\ at lower
redshift just discussed and shown in \Fig\ref{fig:DM}. However, we
caution the reader that this discrepancy can arise from
differences in the sample selection and the analysis of the
datasets, as such as the determination of stellar masses, which
are determined using different apertures for magnitudes, sets of
filters and spectral templates (see \Sec\ref{subsec:systematics}
for further details). We will come back to the dependence on the
redshift in \Sec\ref{sec:DM_evolution}.

We have also compared our results with \TtoSM\ estimates from
gravitational lensing and velocity dispersion of SLACS lenses
(\citealt{Auger+09_SLACSIX,Auger+10_SLACSX}). We have taken lenses
with $\log \mst/\Msun > 11.2$ and an average redshift of $z \sim
0.2$. Lensing data needed to be homogenized in order to be
compared with our \TtoSM\ values in \Fig\ref{fig:DM}, specifically
by: a) converting their size and stellar mass estimates to our
cosmology, b) extrapolating masses from $\Re/2$ to 1 \Re\, and
finally b) de-projecting both stellar and dynamical mass within
\Re. To do that we have adopted for simplicity a SIS model, which
is on average a good approximation of their best-fitted total mass
density, since their fitted power-law models are peaked around an
isothermal slope. Lensing homogenized medians and 25--75th
percentiles are shown with green symbols in \Fig\ref{fig:DM}. An
agreement is found for the \TtoSM--\Re\ and \TtoSM--\rhostar ,
while we notice that at fixed \mst, \sige\ and \Mdyn, SLACS
\TtoSM\ are smaller of $\sim 0.3$ dex than the lower-z KIDS
relation (purple lines). However, at fixed \mst, the SLACS sizes
are smaller than the ones of the KiDS sample by $\sim 0.15$ dex,
while velocity dispersions are higher of $\sim 0.03$ dex, which
implies than that SLACS \Mdyn\ and \TtoSM\ are smaller of $\sim
0.1$ dex within their \Re. The smaller sizes of SLACS galaxies are
also clear from the \TtoSM--\Re\ and \TtoSM--\rhostar\
correlations, where SLACS galaxies have sizes concentrated towards
smaller values, with respect to the range of sizes of SPIDER and
KiDS datasamples. The origin of these discrepancy on sizes of
galaxies of similar stellar mass can reside on the assumption of a
\cite{deVauc48} profile adopted by \cite{Auger+09_SLACSIX} for
their surface photometry fit. It is known that larger Sérsic $n$
values (typical of Massive ETGs) produce \Re s which are larger of
the de Vaucouleurs values of $\sim 0.2$ dex or more
(\citealt{SPIDER-VI}).

In panel (c) of \Fig\ref{fig:DM} we also plot the results from the
\cite{ThomasJ+11}, which make use of Schwarzschild's orbit
superposition models in axisymmetric potentials to a sample of 16
COMA ETGs. We consider their results for a mass-follows-light
model and calculate the \TtoSM\ from their Table~1, dividing the
best-fitted dynamical \ML\ to the Kroupa IMF stellar \ML\
(converted to a Chabrier IMF). Furthermore, for a fair comparison
with our SIS-based results, we have re-scaled their \TtoSM\ using
the average ratio of the virial factors for SIS and constant-\ML\
profile estimated in \cite{SPIDER-VI}. These results are shown as
black dots, and median with 25--75th percentiles are plotted as
black square with error bars. The results are consistent with
SPIDER, but $\sim 0.2$ dex smaller than lower-z KiDS values.
However, as for SLACS lenses, the effective radii adopted by
\cite{ThomasJ+11}, have been obtained fitting a de Vaucouleurs
profile (\citealt{JFK95}; \citealt{Mehlert+00}), which can be
again the reason of the observed discrepancy as their
underestimated \Re\ might have produced smaller \TtoSM .

Our derivation of $\fdm$ yields some cases where galaxies have
unphysical $\fdm < 0$, since $\Mdyn(\Re) < \mst(\Re)$. We find
that only $\sim 6$ per cent of our galaxies have negative DM
fractions. Using a Salpeter IMF we would have obtained smaller DM
fraction, and for $\sim 23$ per cent even negative. We also find
that these numbers are changing with redshift, with $\sim 1$, $3$
and $12$ per cent of negative \fdm\ in the redshift bins $0.1 < z
\leq 0.3$, $0.3 < z \leq 0.5$ and $0.5 < z \leq 0.7$. These
fractions translate to $\sim 5$, $18$ and $34$ if a Salpeter IMF
is adopted. This is a well known critical effect also discussed in
previous works (see e.g. \citealt{Tortora+09}; \citealt{NRT10};
\citealt{SPIDER-VI}). However, although a fraction or almost all
of these negative \fdm\ could be compatible with observational
scatter in \mst\ and \Mdyn\ (see \citealt{NRT10}), it does not
leave a complete freedom on the assumption of the IMF to adopt. In
particular, higher stellar \ML\ normalizations are unphysical for
those systems which tend to have smaller \fdm\ (e.g. the ones with
smaller sizes and dynamical masses, larger stellar densities,
higher redshift, etc.). In principle, one can ask whether by
releasing the concept of the universal IMF it is possible to
interpret all the trends of the mass excess in the central regions
with a stellar mass excess (i.e. an IMF variation) rather than DM
excess (i.e. \fdm\ variation) with galaxy parameters as in
\Fig\ref{fig:DM} (e.g.
\citealt{Tortora+09,SPIDER-VI,TRN13_SPIDER_IMF}; Spiniello's
thesis, Chapter 2).

\subsection{Constraining the IMF}

\begin{figure*}
\centering \psfig{file=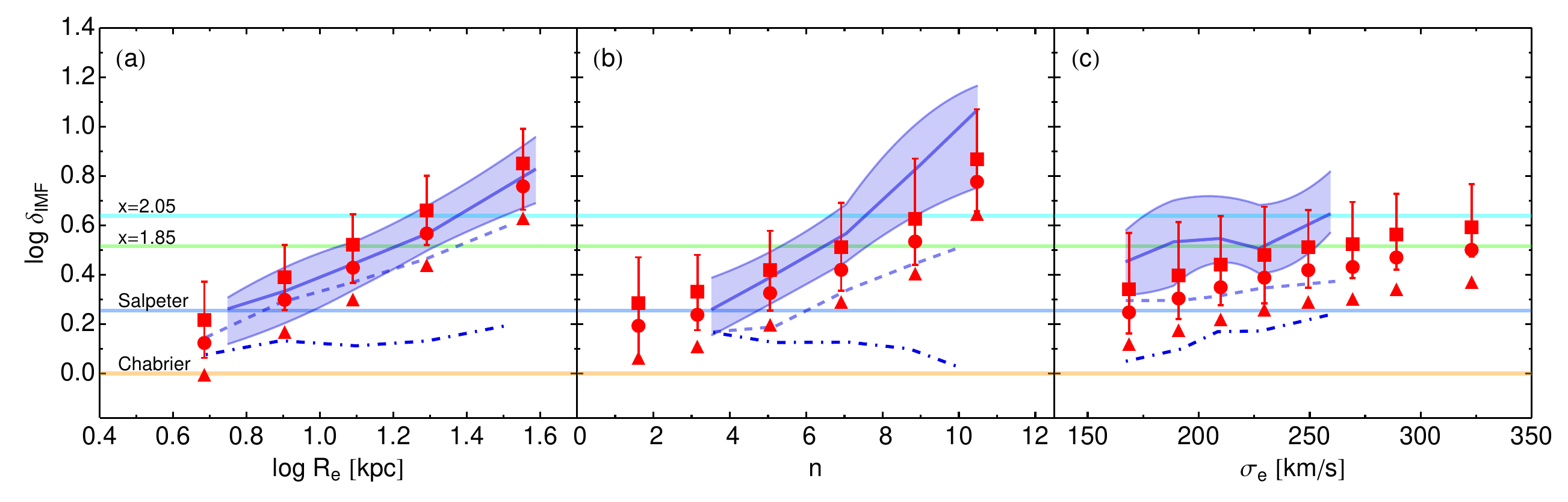,
width=1.00\textwidth} \psfig{file=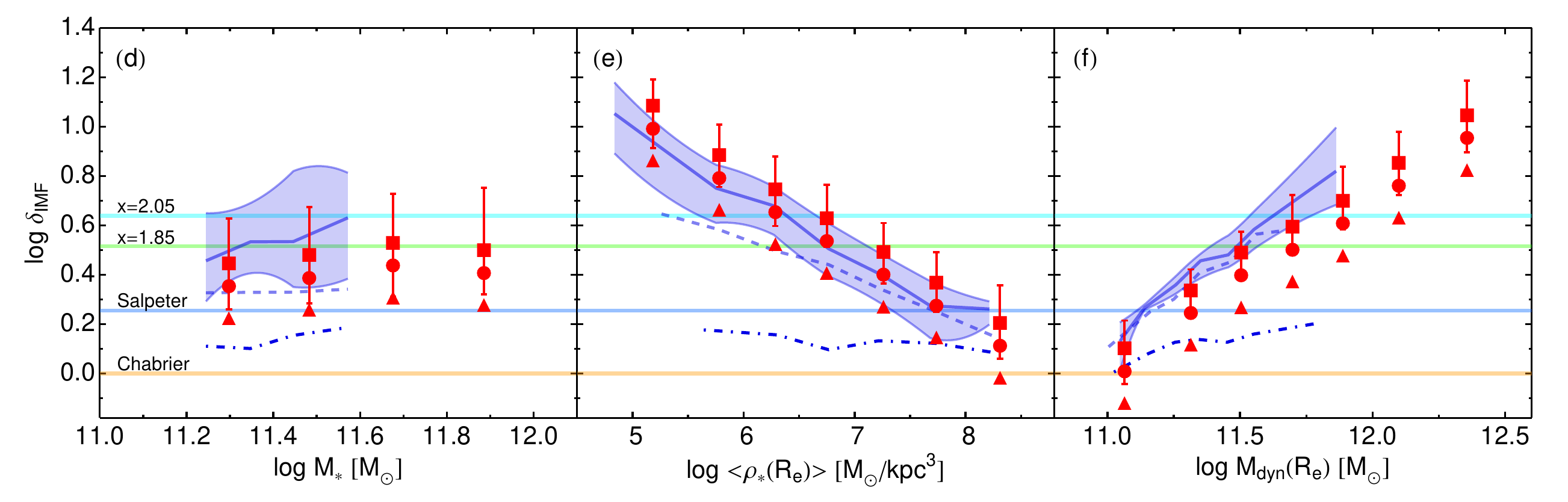,
width=1.00\textwidth} \caption{Trends of the mismatch parameter
\dimf\ as a function of (a) effective radius \Re, (b) S\'ersic
index n, (c) velocity dispersion within effective radius
$\sigma_{\rm e}$, (d) total stellar mass \mst, (e) central average
stellar density \rhostar\ and (f) dynamical mass \Mdyn within \Re,
$\Mdyn(\Re)$. \Re\ and n are rest-frame quantities. The filled red
squares with bars are medians and 25--75th percentiles for a SIS
profile assuming a null DM fraction within \Re. Filled circles and
triangles correspond to the medians adopting a SIS profile with 20
and 40 per cent of DM within \Re, respectively. DM fraction in the
non-universal IMF framework is defined as $\fdm = 1 - M_{\star,\rm
IMF}(\Re)/\Mdyn(\Re)$. Dark blue lines and light blue regions are
medians and 25--75th percentile trends for SPIDER galaxies with
$\mst > 10^{11.2}\, \rm \Msun$. Horizontal lines correspond to the
relative variation of stellar mass, $M_{\star,\rm IMF}/\mst$ --
with respect to a Chabrier IMF -- when adopting synthetic models
with different IMFs, with slopes $1.35$ (i.e. a Salpeter  IMF;
blue), $1.85$ (green), and $2.05$ (cyan). And the orange line
corresponds to the case  of a Chabrier IMF ($M_{\star,\rm
IMF}=\mst$). For completeness and to study systematics in the
trends, we have also plotted the results for the SPIDER sample
using \Ksband\ profile (short-dashed blue line). The point-dashed
blue line is for the NFW + star model with variable stellar \ML\
in \citet{Tortora+14_DMslope}.}\label{fig:IMF}
\end{figure*}

In this section we want to consider the case of a non universal
IMF and correlate the IMF variation with galaxy parameters. The
IMF has been initially considered as universal across galaxy types
and cosmic time, mostly because of a lack of evidence of
variations among stellar clusters and OB associations in the Milky
Way (see \citealt{Bastian+10} for a review about IMF studies).
This assumption has been recently questioned by different
dynamical, lensing, and stellar population studies, finding
evidence for systematic IMF variations in ETGs (\citealt{Treu+10};
\citealt{ThomasJ+11}; \citealt{Conroy_vanDokkum12b};
\citealt{Cappellari+12, Cappellari+13_ATLAS3D_XX};
\citealt{Spiniello+12}; \citealt{Wegner+12}; \citealt{Dutton+13};
\citealt{Ferreras+13}; \citealt{Goudfrooij_Kruijssen13};
\citealt{LaBarbera+13_SPIDERVIII_IMF}; \citealt{TRN13_SPIDER_IMF};
\citealt{Weidner+13_giant_ell}; \citealt{Goudfrooij_Kruijssen14};
\citealt{Shu+15_SLACSXII};
\citealt{Tortora+14_DMslope,Tortora+14_MOND};
\citealt{Martin-Navarro+15_IMF_variation}; \citealt{Lyubenova+16};
\citealt{TLBN16_IMF_dwarfs}; \citealt{Corsini+17};
\citealt{Li+17_IMF}; \citealt{Sonnenfeld+17_IMF}).

Following \cite{SPIDER-VI} and motivated by all the results
pointing to a non-universal IMF (see in particular
\citealt{Tortora+14_DMslope}), we can check how large the IMF has
to be to cancel the presence of DM within 1 \Re. We estimate the
expected variation of stellar mass normalization, defining the
mismatch parameter $\dimf \equiv M_{\star,\rm
IMF}(\Re)/\mst(\Re)$, relative to a Chabrier IMF, where \mst\ is
the stellar mass estimated with a Chabrier IMF and $M_{\star,\rm
IMF}$ is the stellar mass for any other IMF assumption. We can
also define the related DM fraction as the one obtained for the
particular IMF assumed, i.e. $\fdm = 1 - M_{\star,\rm
IMF}(\Re)/\Mdyn(\Re)$. The estimated \dimf\ with the assumption
$\fdm=0$ is substantially the \Mdyn/\mst\ plotted in
\Fig\ref{fig:DM}, but with $\Mdyn = M_{\star,\rm IMF}$. The IMF
mismatch parameter corresponding to $\fdm=0$ represents an extreme
upper limit, since in the $\Lambda$CDM a non-zero DM fraction is
found also when accounting for a non-universal IMF (see e.g.
\citealt{Cappellari+13_ATLAS3D_XX}; \citealt{TRN13_SPIDER_IMF}).
Hence, to explore more realistic dark matter fractions, we have
also computed the IMF mismatch corresponding to \fdm$=0.2$ or
\fdm$=0.4$, which bracket the typical values of the DM fraction
found in SPIDER ETGs, when a free IMF normalization is adopted
(\citealt{TRN13_SPIDER_IMF}). By construction, for the $\fdm =
0.2$ and $0.4$ cases, our \TtoSM\ give a mass budget in stars, and
thus \dimf\ values, which are systematically lower than the $\fdm
= 0$ case.

To derive inferences about the IMF slope, we compare this
dynamical \dimf\ with what is expected from stellar population
models. Thus, we consider three power-law IMFs, with slopes $1.35$
(i.e. Salpeter), $1.85$, and $2.05$ (i.e. a very bottom-heavy
IMF). The \dimf\ is estimated as the ratio of the stellar \ML\
between two SSPs having a power-law and a Chabrier IMFs,
respectively. We compute the stellar K-band $M/L$, adopting the
\cite{BC03} synthesis code, for old ($10$~Gyr) SSPs, with solar
metallicity (see \citealt{SPIDER-VI}).

\Fig\ref{fig:IMF} plots the \dimf\ trends, for three different
assumptions of the DM fraction, as a function of \Re, n, \sige,
\mst, \rhostar\ and \Mdyn\ (red symbols). We compare the results
against the $z\sim0$ estimate from SPIDER for \fdm$=0$. Horizontal
lines mark \dimf\ for the different IMFs obtained from stellar
population synthesis. The intersections with the horizontal lines
define the values of \Re, n, \sige, \mst, \rhostar\ and \Mdyn\,
for which a given IMF slope would imply \fdm$=0$, $0.2$ and $0.4$.

The figure shows that, in order to account for the apparent trend
of \fdm\ with \Re, n, \rhostar\ and \Mdyn, galaxies with the
largest radii, S\'ersic indices and dynamical masses and the
lowest \rhostar\ should have an IMF slope as steep as (or steeper
than) $2.05$ (e.g. \citealt{TRN13_SPIDER_IMF};
\citealt{Tortora+14_DMslope};
\citealt{Spiniello+15_IMF_vs_density}). While at the lowest values
of \Re, n and \Mdyn, and highest \rhostar, a Salpeter (Chabrier)
IMF would be required if \fdm$=0$ (\fdm$=0.4$). Interestingly,  in
our mass range, \dimf\ is almost constant with stellar mass and is
consistent with a slope $\sim 1.8$ when \fdm$=0$, and with a
Salpeter IMF if \fdm$=0.4$. Despite the adopted constant \fdm, the
trends with the parameters remain qualitatively the same and just
shifted toward a lower normalization for higher \fdm\ values.

However, IMFs with slopes $x\gsim \, 2$ are disfavored in SLACS
gravitational lenses, since they give stellar \ML s which violate the
total mass inside the Einstein radius, i.e. within $\sim \Re/2$
(\citealt{Spiniello+12}). Thus, the assumption of no DM within \Re\
is not realistic and would be at odds with other results using
gravitational lensing, too (see also later on
in the paper).

Before starting to drive conclusions, we need to check how our
assumptions might affect the main results of our analysis. All
sources of systematics will be discussed in
\Sec\ref{subsec:systematics}, however here we start showing the
effect of unaccounted colour gradients and the galaxy model, using
the SPIDER sample. For this sample we show the K-band results as
dashed blue lines in \Fig\ref{fig:IMF}, which provide smaller
\dimf\ of about $0.2$ dex, but with trends that are almost
unchanged. We also plot the best fitted \dimf\ derived in
\cite{Tortora+14_DMslope}, using a standard NFW for the DM halo
and a K-band light profile (dot-dashed blue line). The differences
among these two results are naturally explained by the fact that
dashed blue lines assume a SIS profile for the total mass
distribution and $\fdm = 0$, while point-dashed blue lines
correspond to a NFW plus light model, which also predict non-zero
DM fractions (\citealt{TRN13_SPIDER_IMF};
\citealt{Tortora+14_DMslope}).

\section{Evolution with redshift}\label{sec:DM_evolution}

A simple monolithic-like scenario, where the bulk of the stars is
formed in a single dissipative event followed by a passive
evolution, is not longer supported by the observations, while many
observations suggest the occurrence of a strong mass and size
evolution in ETGs (\citealt{Daddi+05}; \citealt{Trujillo+06};
\citealt{Trujillo+07}; \citealt{Saglia+10}; \citealt{Trujillo+11};
\citealt{Tortora+14_DMevol}). In this section, we will first
investigate the evolution of size and DM fraction as a function of
redshift, at fixed stellar mass, comparing the results with some
literature and predictions from different galaxy evolution
scenarios. Then, to study in more detail the evolution of the mass
and size in our galaxy sample, we compare some relevant
correlations, as the ones between the galaxy size or DM fraction
and \mst, at different redshifts (\citealt{Tortora+14_DMevol}). In
this latter case, we will create some toy-models to interpret this
evolution in terms of physical processes.

Previous analyses addressing the DM fraction evolution with
redshift (e.g. \citealt{Tortora+14_DMevol} based on the EDisCS
sample and \citealt{Beifiori+14} based on BOSS) have shown that
ETGs contain less DM within the effective radius at larger
redshift. In this paper, we will complement our analysis in
\cite{Tortora+14_DMevol}, determining the DM evolution
self-consistently, i.e., using the same datasample processed with
exactly the same analysis.

\subsection{Evolution at fixed mass}\label{subsec:evol_fix_mass}

\begin{figure*}
\centering
\includegraphics[trim= 0mm 0mm 0mm 0mm, width=0.705\textwidth,clip]{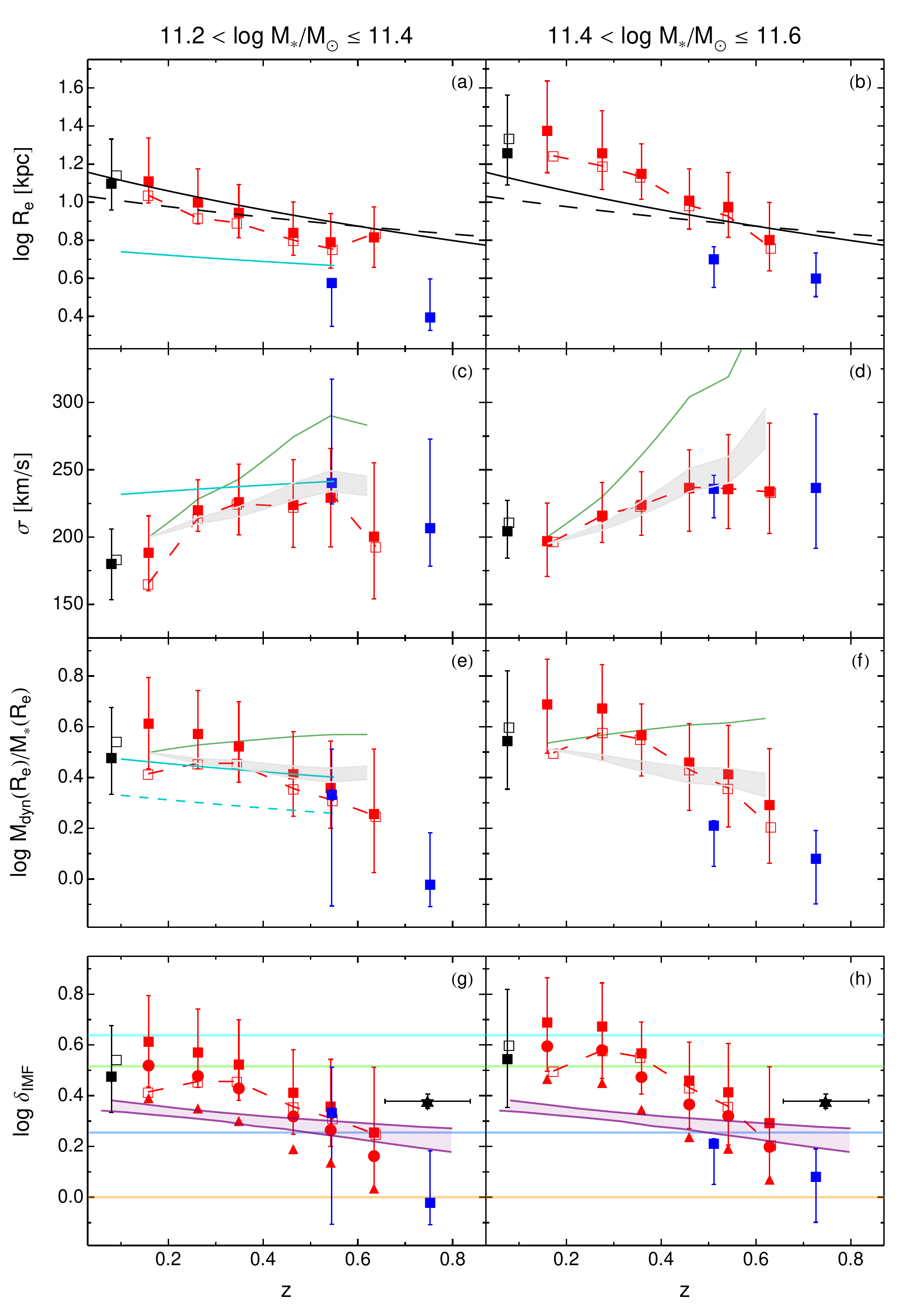}
\includegraphics[trim= 0mm -9mm 5mm 0mm, width=0.285\textwidth,clip]{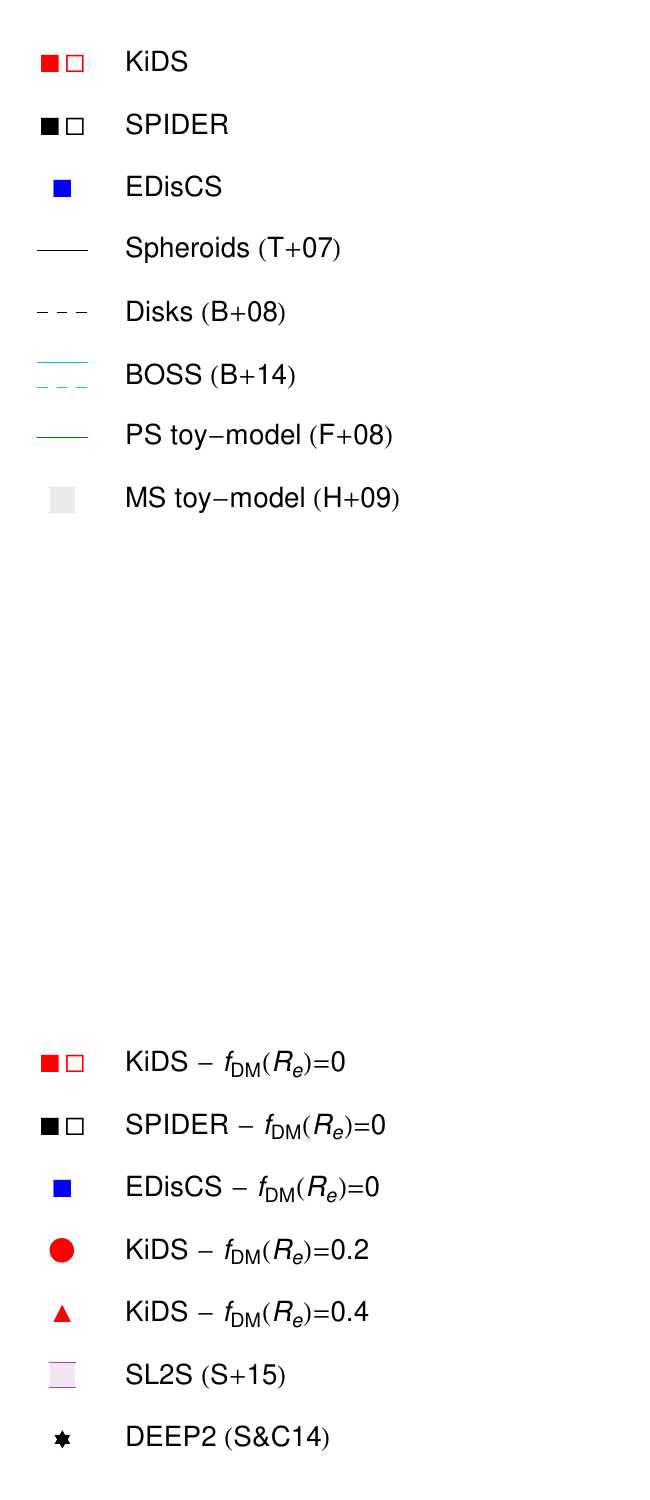}
\caption{Evolution with redshift of \Re\ (panels a and b), \sige\
(panels c and d), \TtoSM\ (panels e and f) and \dimf\ (panels g
and h) for two stellar mass bins: $11.2 < \log \mst/\Msun \leq
11.4$ (left) and $11.4 < \log \mst/\Msun \leq 11.6$ (right). Red
symbols are as in previous figures. Open black square with error
bar is median and 25--75th percentiles for SPIDER galaxies, while
open black square is the median for the SPIDER sample when the
progenitor bias is taken into account. Blue squares with bars are
median and 25--75th percentiles for EDisCS sample. Cyan solid
lines in panels (a), (c) and (e) are from
\citet[B+14]{Beifiori+14}. In panel (e) the dashed line is
calculated converting the results in \citet{Beifiori+14} assuming
a constant-\ML\ profile to a SIS profile, as explained in the
text. The black solid and dashed lines in the top panels are taken
from the average \Re/$R_{\rm e,0}$--z trends (with $R_{\rm e,0} =
\Re(z\sim 0)$) for spheroid-like galaxies in
\citet[T+07]{Trujillo+07} and disk-like systems in
\citet[B+08]{Buitrago+08}, normalized to $R_{\rm e,0} = 15$ and
$11$ kpc, respectively. Shaded gray region and green line are our
predictions using the merging model of
\citet[H+09]{Hopkins+09_DELGN_IV} and the "puffing-up" scenario
from \citet[F+08]{Fan+08}, respectively. See the text for further
details. In the bottom panels (g and h), we show the \dimf\ when
$\fdm=0$ (filled squares) $\fdm = 0.2$ (filled circles) and $\fdm
= 0.4$ (filled triangles), as in \Fig\ref{fig:IMF}. Purple lines
and shaded regions are the results obtained from the strong
lensing and dynamical analysis of 80 ETGs in
\citet[S+15]{Sonnenfeld+15_SL2SV} and \citet{Sonnenfeld+17_IMF},
the region sets the 68 per cent confidence level. The black star
with error bars is for the average \dimf\ obtained by analyzing
the kinematical data of a sample of 68 galaxies at $z\sim 0.75$ in
\citet[S\&C14]{Shetty_Cappellari14}. Horizontal lines correspond
to the relative variation of stellar mass, $M_{\star,\rm
IMF}/\mst$ -- with respect to a Chabrier IMF -- as in
\Fig\ref{fig:IMF}. See legend at right, explanation in the main
text, and the text in this caption for
abbreviations.}\label{fig:evolution}
\end{figure*}

Following some previous studies about size and velocity dispersion
evolution we investigate how \Re, \sige\ and \TtoSM\ change in
terms of redshift, at fixed stellar mass. We concentrate our
attention on two particular stellar mass bins ($11.2 < \log
\mst/\Msun \leq 11.4$ and $11.4 < \log \mst/\Msun \leq 11.6$).
Almost all the correlations discussed are significant at more than
99 per cent. In \Fig\ref{fig:evolution} we first plot the
dependence of \Re\ with the redshift (panels a and b), which show
that sizes were smaller at earlier epochs of galaxy evolution (see
Roy et al. in preparation, for further details about size
evolution in KiDS galaxies). Following a standard approach in the
literature, we fit the relation $\Re = R_{\rm e,0}(1+z)^{\alpha}$
to the data. For the two mass bins, in the case of no progenitor
bias correction (red filled squares with error bars), we find a
slope, $\alpha = -2.2$ and $-3.8$ respectively. These values
translates into a weaker size evolution if we account for the
progenitor bias (open squares with dashed red lines): in fact, the
slopes become $\alpha = -1.6$ and $-3.3$ in this case. These
trends are steeper than the trends for spheroid- and disk-like
systems with $\mst > 10^{11}\, \rm \Msun$ taken from the
literature (solid and dashed black lines in the top panels in
\Fig\ref{fig:evolution}; \citealt{Trujillo+07};
\citealt{Buitrago+08}; \citealt{Conselice14_review}). At lower z
we find a good agreement with the \Re s from SPIDER datasample.
However, we find some discrepancy in the size normalization with
other analysis. For example, the I-band measurements from EDisCS
sample (\citealt{Saglia+00}; \citealt{Tortora+14_DMevol}) are
lower of $\sim 0.3$ dex, while the $i$-band \Re\ in
\cite{Beifiori+14}\footnote{Note that the mass range used by
\cite{Beifiori+14} is not exactly the same of the first mass bin,
used in this paper.}, re-calibrated using HST images, are smaller
of a factor $\sim 0.2$ dex. The difference in the wavebands
adopted (\iband\ \Re\ in \cite{Beifiori+14} vs our rest-frame \Re)
cannot account for the observed large discrepancy. Overall, our
results confirm the well known result that in massive galaxies the
size of the galaxies is changing with redshift
(\citealt{Daddi+05}; \citealt{Trujillo+06}; \citealt{Trujillo+07};
\citealt{Buitrago+08}; \citealt{vanderWel+08}).

In panels (c) and (d) of \Fig\ref{fig:evolution} we also plot the
effective velocity dispersion, \sige, as a function of the
redshift. In this case the evolution with redshift is shallower,
with higher--z galaxies having slightly larger velocity
dispersions (\citealt{Cenarro_Trujillo09}; \citealt{Posti+14}). In
this case the evolution of \sige\ is quantified through the
relation $\sige = \sige_{\rm ,0}(1+z)^{\alpha}$ where the
estimated slopes for the two mass bins above are $\alpha = 0.21$
and $0.53$ (without progenitor bias) and $\alpha = 0.46$ and
$0.53$ (with progenitor bias) respectively. These results are in
good agreement with local (\citealt{SPIDER-I};
\citealt{SPIDER-VI}), intermediate-z (\citealt{Beifiori+14}) and
higher-z (\citealt{Saglia+10}; \citealt{Tortora+14_DMevol})
measures.

The total-to-stellar mass ratio (with a Chabrier IMF) is plotted
in panels (e) and (f) of \Fig\ref{fig:evolution}. The galaxies are
DM dominated at lower redshift (i.e. 75--80 per cent of DM at $z
\sim 0.2$), while \fdm\ results to be smaller at higher--z (40--50
per cent at $z \sim 0.6$). Fitting the $\TtoSM = (\TtoSM)_{\rm
0}(1+z)^{\alpha}$ relation to the data, for the two mass bins we
find $\alpha = -2.4$ and $-2.8$ (without progenitor bias) and
$\alpha = -1.3$ and $-2.2$ (with progenitor bias) respectively. We
find a small discrepancy with SPIDER and EDisCS data sample, but
these results agree within the data scatter. We also plot the
\TtoSM\ derived in \cite{Beifiori+14}, assuming a non-homologous
constant-\ML\ profile (with SDSS sizes re-calibrated to HST
values) as a dashed cyan line. This latter model cannot be
directly compared to our results because of its different
assumption on the total mass distribution, hence we have re-scaled
their \TtoSM\ using the average ratio of the virial factors for
SIS and constant-\ML\ profile estimated in \cite{SPIDER-VI}. After
this re-normalization (solid cyan line in \Fig\ref{fig:evolution})
the \cite{Beifiori+14} estimates are on average consistent within
$1 \, \rm \sigma$ scatter with the KiDS sample.

We want to interpret the trends of \sige\ and \TtoSM\ in the
context of galaxy evolution, by comparing the observed trends with
the predictions from two different scenarios invoked to explain
the galaxy size evolution. The merging scenario (MS, hereafter)
predict that size is driven by the accretion of matter, with the
result that the sizes of the merger remnants are larger than those
of their remnants. The merging model of \cite{Hopkins+09_DELGN_IV}
predicts that the velocity dispersion varies in terms of the size
as $\sigs(z) \propto (1+\gamma)^{-1/2} \sqrt{\gamma +
\Re(0)/\Re(z)}$, where the parameter $\gamma$ sets the DM
contribution to the potential relative to that of the baryonic
mass. This parameter varies between 1 and 2 (which are the best
fitted values for $\mst \sim 10^{11}$ and $\sim 10^{12}\, \rm
\Msun$, respectively). Completely different is the "puffing-up"
scenario (PS, hereafter) from \cite{Fan+08}, which predict that
galaxies grow by the effect of quasar feedback, which removes huge
amounts of cold gas from the central regions, quenching the star
formation and increasing the size of the galaxy. This model
predicts that velocity dispersion varies as $\Re^{-1/2}$.

To derive predictions in the above scenarios, we use as \Re--z
relation the interpolating line going through the KiDS median
values in panels (a) and (b) of \Fig\ref{fig:evolution}. This
latter is inserted into the two equations discussed to derive the
predicted velocity dispersions in the two schemes. Then we need to
translate these predicted velocity dispersions into a \TtoSM. In
order to do this we need first to derive the \Mdyn\ as a function
of the redshift, solving the spherical Jeans equation, which
contains 1) the density of the light distribution, 2) the total
potential, 3) all as a function of redshift. For the light
distribution we have taken the S\'ersic profile with $n=4$ for
simplicity (i.e. a pure de Vaucouleurs) and with effective radius
given by our interpolated $\Re(z)$ relation as defined above. For
the total potential we have used the SIS profile. Then, we impose
that the velocity dispersion derived from Jeans equation (averaged
within \Re) equals the $\sigma(z)$ in the two scenarios, MS and
PS. This procedure provides \Mdyn\ and \TtoSM\ as a function of
redshift. Note that what is relevant in this calculation is the
trend with redshift and not the normalization, which is fixed by
hand, since in the $\sigma_{\star}(z)$ formulae the normalization
factor is unspecified.

We plot the predicted trends for \sige\ and \TtoSM\ in panels (c),
(d), (e) and (f) of \Fig\ref{fig:evolution}. The PS predicts a
very strong evolution (with a change of $\sim 100\, \rm km s^{-1}$
in the redshift window analyzed), which is discrepant with KiDS
results for both the \sige\ and \TtoSM. On the contrary, the
milder evolution from MS accommodates the observations. See a
similar analysis for the $\sigma$ evolution in
\cite{Cenarro_Trujillo09}, where similar conclusions are reached.
However, while the agreement with the velocity dispersion seem
very good, the model predict a shallower \TtoSM--z trend than the
observed one, in both the mass bins.

Following the analysis made in \Sec\ref{sec:DM_correlations}, in
panels (g) and (h) we also interpret the evolution of the \TtoSM\
with redshift in terms of a change in the IMF and of DM content.
We show how the mismatch parameter \dimf\ is changing in terms of
redshift, for 3 different \fdm\ values ($\fdm = 0$, $0.2$ and
$0.4$). If we assume that $\fdm(\Re)$ is not changing with
redshift, then the observed \TtoSM\ evolution can be explained by
a change of the IMF "normalization". The \dimf\ values found for
$\fdm=0$ need to be considered as upper limits, and point to a
Salpeter IMF at $z \sim 0.65$ and a very steep IMF at low redshift
(with $x \sim 2.05$), if $\fdm = 0$. For the maximal DM fraction
adopted here, i.e. $\fdm = 0.4$, KiDS results suggest a standard
Chabrier IMF at $z \sim 0.65$ and slightly super-Salpeter at
$z\sim 0$. If we consider that a more realistic situation would
allow for a variation of \fdm\ with redshift (e.g., from $\fdm =
0$ at $z \sim 0.65$ to $\fdm = 0.4$ at $z\sim0$), then the \dimf\
evolution would result weaker or totally absent.

We finally compare our findings for \dimf\ with some results from
the literature. \cite{Sonnenfeld+15_SL2SV} and
\cite{Sonnenfeld+17_IMF} measure \dimf\ for a sample of 80 massive
ETGs, using strong lensing and velocity dispersion data. We plot
their 68 per cent confidence region with purple symbols in
\Fig\ref{fig:evolution}. Their trend is shallower than our results
found assuming a constant \fdm\ value. An agreement with these
results can be found if we allow in our data for the more
realistic \fdm\ evolution discussed above, i.e. if we assume that
\fdm\ is null at $z \sim 0.65$ and $\gsim \, 0.4$ at $z \sim 0$.
The black star with error bars is the average \dimf\ from
\cite{Shetty_Cappellari14}, obtained through the dynamical
modelling of kinematical data of a sample of 68 massive ($\mst
\gsim 10^{11}\, \Msun$) galaxies at $z\sim 0.75$, extracted from
the DEEP2 spectrographic survey. As in our $\fdm = 0$ case, their
results have to be considered as upper limits, since they assume
that mass follows light and no DM at all is considered. Their
estimate would be normalized to our mass model assumption. This
would be done by adding a factor $\sim 0.14$ dex to covert a
constant \ML\ profile to a SIS (\citealt{SPIDER-VI}), but their
value would be larger of $\sim 0.16$ dex if compared with the
virial estimator, which they suggest to be related to
underestimated \Re\ in the virial estimator formula (see
\citealt{Cappellari+13_ATLAS3D_XX};
\citealt{Shetty_Cappellari14}). Thus, since these two factors
would almost elide, we have decided to not add any corrective
factor to their \dimf\ estimate shown in \Fig\ref{fig:evolution}.
These results are about $0.1$ dex larger than our $\fdm=0$
findings at the same redshifts (see KiDS and EDisCS data points in
\Fig\ref{fig:evolution}).

As it has been done for the central DM evolution presented in
panels (e) and (f) in \Fig\ref{fig:evolution}, it would be
interesting to interpret if this change of IMF with time is
realistic and if is consistent with specific processes. In
\cite{Tortora+14_DMevol} we proposed some speculative
considerations, suggesting that a change of IMF with redshift
could be produced by two different processes: (a) new stars formed
in the galaxy center during a wet merging process, which also
produce 'higher mass' IMF (\citealt{NRT10}) and positive age
gradients (\citealt{Tortora+10CG}) in young and massive local
ETGs, or (b) by stars from both merging galaxies, which are
characterized by two different IMFs, which can combine to modify
the cumulative IMF of the merger remnant. Unfortunately, the net
effect of these processes on the final IMF normalization is not
yet clear, and in most cases the combination of a 'higher' and a
'lower mass' IMF would produce a diluted IMF. Thus, galaxy mergers
would produce smaller \dimf\ values, or leave \dimf\ almost
constant with redshift, as predicted by the galaxy merging
toy-models in \cite{Sonnenfeld+17_IMF}. These results contrast the
strong evolution observed in panels (g) and (h) of
\Fig\ref{fig:evolution}.

All these results seem to suggest that most of the \TtoSM\
evolution would be driven by a change of the DM fraction. A deeper
analysis will be needed to constrain the IMF and DM fraction in
term of redshift. In a future paper we will discuss this problem
with more details.

\subsection{Size, mass and DM evolution}\label{subsec:evol_trends}

\begin{figure*}
\centering \psfig{file=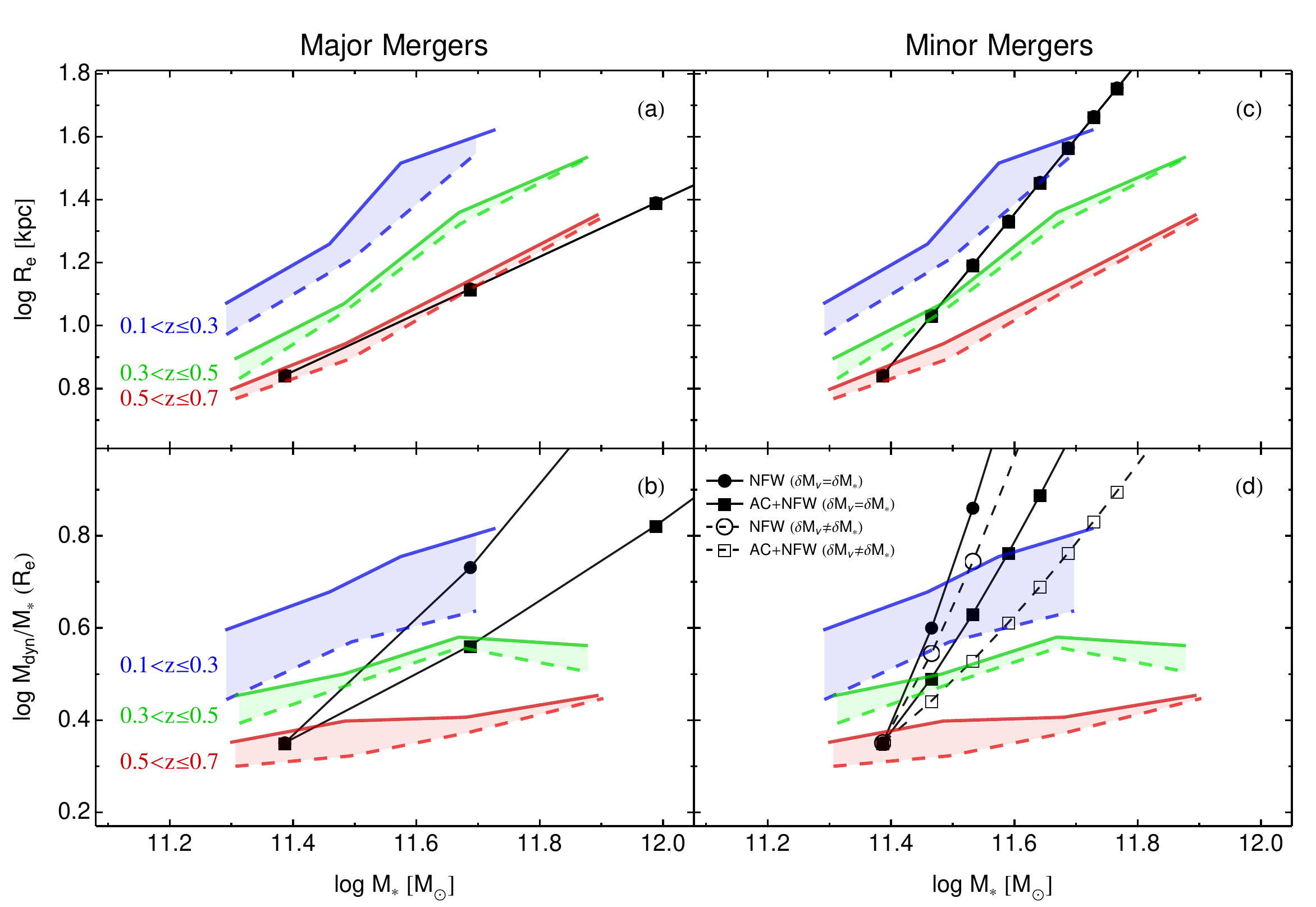,
width=0.80\textwidth} \caption{Evolution of \Re--\mst\ (panel a
and c) and \TtoSM--\mst\ (panels b and d). Blue, green and red
lines are for galaxies in three redshift bins $0.1 < z \leq 0.3$,
$0.3 < z \leq 0.5$ and $0.5 < z \leq 0.7$. Solid lines are medians
for the full sample, while dashed lines are for results corrected
for progenitor bias. We take, as example, the average galaxy at
$\log \mst/\Msun = 11.4$ and evolve it accordingly to the
toy-models discussed in the text. In the left (right) panels major
(minor) merging predictions are shown. In the bottom panels (b and
d), dots and filled squares are for NFW and AC+NFW profiles when
$\delta \mst = \delta \Mvir$, and set the \Re, \mst\ and \TtoSM\
after each single merging event. In panel d, open circle and
squares are for NFW and AC+NFW profiles when $\delta \mst \neq
\delta \Mvir$ }\label{fig:evol_toy_models}
\end{figure*}

The analysis performed in the previous section cannot be
conclusive, since it does not take into account that in merging
processes a single galaxy also changes in mass. For this reason,
in \Fig\ref{fig:evol_toy_models} we compare the \Re--\mst\ and
\TtoSM--\mst\ relations (assuming a fixed Chabrier IMF) for KiDS
galaxies at different redshift and study the joint evolution of
size and mass. We start seeing how lower redshift galaxies are
larger and contain more DM in their cores at all values of \mst,
confirming the trends in \Figs\ref{fig:DM} and
\ref{fig:evolution}. The trend is weaker if we consider the
progenitor bias, which affects mostly the lowest redshift bin
(dashed lines in \Figs\ref{fig:DM} and \ref{fig:evolution}). In
the following we will try to reproduce these trends within the
hierarchical framework. Galaxy mergers are the most accredited
mechanisms that can account for both size and mass accretion, as
we have also seen from the analysis of
\Sec\ref{subsec:evol_fix_mass}. Dissipationless major mergers from
simulations in elliptical galaxies have predicted that the DM
fraction within a certain physical radius decreases mildly after
the merger (\citealt{Boylan-Kolchin+05}). But they have also shown
that the DM fraction within the final \Re\ is greater than the DM
fraction within the initial \Re, because the total mass within
\Re, $M_{\rm tot}(\Re)$, changes after the merger more than
$\mst(\Re)$. We have found the same result with real data
(\Fig\ref{fig:evolution}; \citealt{Tortora+14_DMevol}) and we have
also found in \Fig\ref{fig:evolution} that this is also the case
for a toy-model which assumes the merging model from
\cite{Hopkins+09_DELGN_IV} and the observed \Re--z relation. More
recently, the problem has been investigated in detail with N-body
simulations by \cite{Hilz+13}. At different final stellar masses,
they find that the equal-mass mergers produce a smaller size
increase of multiple minor mergers. In particular, the variation
of \Re\ with respect to the initial radius, $\Re / R_{\rm 0}$, in
terms of the variation of \mst\ with respect to the initial
stellar mass, $\mst / M_{\rm 0}$, is found to be $\Re/R_{\rm 0}
\propto$ $(\mst/M_{\rm 0})^{0.91}$ for the equal-mass merger and
$\propto (\mst/M_{\rm 0})^{2.4}$ for the minor mergers
(consistently with what was observed by \citealt{vanDokkum+10}).

Taking these results into account, we have constructed some
toy-models assuming that $M_{\rm DM} \propto \Mvir r^{\eta}$
around \Re, with $\eta \sim 2$ for a standard NFW and $\eta \sim
1.2$ for a contracted NFW, hereafter AC+NFW (according with
\citealt{Boylan-Kolchin+05}). Following \cite{Hilz+13}, we have
also taken the average evolution of \Re\ in terms of \mst\
evolution for the equal-mass merging (i.e. $\Re/R_{\rm 0} \propto
(\mst/M_{\rm 0})^{0.91}$) and minor merging (i.e. $\Re/R_{\rm 0}
\propto (\mst/M_{\rm 0})^{2.4}$). Following
\cite{Tortora+14_DMevol}, we start assuming that the variation of
the virial mass follows the one of the stellar mass, i.e. $\delta
\Mvir \approx \delta \mst$. This intrinsically reflects the
hypothesis that the systems participating to merging (i.e. the
progenitors) all have the same $\Mvir / \mst$, which is reasonable
for most of the stellar mass range covered by our sample. However,
it is possible - mainly for the minor merging case - that the
virial mass change at a different rate of the stellar mass, since
the main galaxy is merging with another galaxy with a different
$\Mvir/\mst$ (see, e.g., Table 2 in \citealt{Boylan-Kolchin+05},
where the final virial mass after the mergers is changed less than
total stellar mass). These simplified models provide a
quantitative assessment of the impact of the merging scenario on
the observed DM fraction.

We have considered the evolution tracks related to the two
different merging types (major and minor) for a galaxy with
$\log\mst/\Msun = 11.4$. Dots and squares are for NFW and AC+NFW
profiles, respectively. In the left panels (a and b), the major
merging tracks are shown as black lines with dots/squares
indicating the events corresponding to masses $\delta \mst \times
M_{\rm \star, 0}$, with the mass increments $\delta \mst =
1,2,4,...$, assuming that $\delta \Mvir = \delta \mst$. The first
dot/square at $\log \mst/\Msun = 11.4$ corresponds to the initial
progenitor galaxy. The second dot/square is the result of one
major merger, which doubles the initial mass of the progenitor,
while the third dot/square corresponds to a second major merger
with mass 4 times the mass of the initial progenitor and 2 times
the mass of the remnant of the first merging event. The minor
merging tracks are shown in the right panels (c and d) by black
lines. Dots/squares indicate remnant with masses $M_{\rm \star, 0}
+ \delta \mst \times M_{\rm \star, 0}$ where $\delta \mst
=0,0.2,0.4,0.6,...$, and we use two different increment laws for
\Mvir. In panel (b) we use $\delta \Mvir = \delta \mst$. In panel
(d) filled symbols are for $\delta \Mvir = \delta \mst$, while
open symbols correspond to $\delta \Mvir = 0.05 \delta \mst$. In
this case we suppose that the initial progenitor galaxy with mass
$M_{\rm \star, 0}$ is experiencing a collection of minor mergers
with galaxies having masses of $20$ per cent of $M_{\rm \star,
0}$.

In order to determine which type of merging is describing the mass
assembly of these massive galaxies we compare the model
predictions with the data in \Fig\ref{fig:evol_toy_models}. We
plot the data without (with) progenitor bias as solid (dashed)
lines. Blue, green and red lines are for galaxies in three
redshift bins $0.1 < z \leq 0.3$, $0.3 < z \leq 0.5$ and $0.5 < z
\leq 0.7$. We aim at whether mass, size and total-to-stellar mass
evolution in KiDS galaxies can be explained, consistently, through
major or minor mergers. We see that major mergers can be excluded,
since the predicted evolution in size of a galaxy in the highest
redshift bin (with $z \sim 0.6$) and with stellar mass
$10^{11.4}\, \Msun$ is parallel to the size-mass relation in this
same redshift bin (see left panels in
\Fig\ref{fig:evol_toy_models}). On the contrary, the same galaxy
can evolve to $z \sim 0.2$ experiencing few (5 or 6) minor
mergers, which accrete $\sim 100$ per cent of the initial stellar
mass $M_{\rm \star, 0}$.

The example galaxy is not evolving on the top of the $z \sim 0.2$
\Re--\mst\ and \TtoSM--\mst\ relations if we consider that DM and
star accrete at the same rate, for both NFW and AC+NFW (dots and
filled squares in \Fig\ref{fig:evol_toy_models}). The
\TtoSM--\mst\ evolution is too steep, thus after $\sim 2$ minor
mergers the example galaxy would end up on the $z \sim 0.2$
observed \TtoSM--\mst, this number is not consistent with what
found analyzing the \Re--\mst\ evolution, which would require
$4-6$ mergers to transform the $z \sim 0.6$ galaxy in a typical
galaxy observed at $z \sim 0.2$. The number of minor mergers
needed to transform the $z \sim 0.6$ galaxy in a bigger and more
DM-dominated galaxy is found if the DM mass is accreting with a
lower rate and only for the AC+NFW mass profile. This is possible
if the main progenitor is merging with lower mass galaxies with
smaller \TtoSM. This result suggest that our massive galaxy at $z
\sim 0.6$ has to merge with a population of less massive galaxies
with lower total-to-stellar mass ratios. Abundance matching
studies predict the existence of these systems, since at stellar
masses smaller than $10^{11.4} \, \Msun$ galaxies are found to
have, on average, smaller \TtoSM\ (see e.g. \citealt{Moster+10}).

\subsection{Systematics}\label{subsec:systematics}

There are different sources of systematics that may impact our
results. We will list some of these systematics in the following

\begin{itemize}
\item[(i)] {\it KiDS vs. SPIDER.} We have cross-matched the full matched DR7 sample (5223 galaxies) with the
SDSS-based SPIDER datasample (39993 galaxies), finding 248
galaxies with reliable structural parameters and masses in common.
This sample spans the range of masses $\log \mst/\Msun \sim
(10.2-11.4)$, which is different from the range used in this paper
(i.e., $\log \mst/\Msun > 11.2$). We compare both structural
parameters, stellar masses and velocity dispersions. Some
differences are expected since the two samples rely on different
image quality. Thus, while the median axis ratio is identical, and
equal to $q_{\rm r} = 0.7$, some small differences, are found
between KiDS and SPIDER in the effective radius ($3.2''$ vs.
$2.9''$) and S\'ersic index (5.6 vs. 5.3). The reason of this
difference can probably be examined in the deepest KiDS images,
which, including more light from the external regions, tend to
produce larger sizes. However, these differences are within the
typical measured uncertainties (see Roy et al. 2017, in
preparation, for further details). Moreover, also stellar masses
are consistent, with a median value of $\log \mst \Msun \sim 10.7$
dex. But, we caution the reader that KiDS and SDSS stellar masses
are calculated using different photometric apertures (3 arcsec vs
3 times the $i$-band Kron radius), different wavelength coverage
(optical vs. optical+NIR) different stellar population
prescription (single-bursts vs. exponential star formation). The
aperture velocity dispersion is also consistent.

\item[(ii)] {\it Stellar mass modelling.} There are many
systematics related to the stellar mass modelling. First of all,
the \mst\ (\TtoSM) can become lower (higher) if we relax the
constraints on age (e.g. by allowing ages lower than $3\, \rm
Gyr$) and metallicity (hence resuming the full BC03 synthetic
library). The trends in terms of redshift are shallower, since the
masses at larger redshifts are underestimated, as discussed in
\cite{Maraston+13_BOSS}. Then, we have seen that median masses
from \textsc{le phare} are smaller or $\sim 0.05$ dex if compared
with the best-fitted values adopted in our analysis; however, the
trends with redshift for \TtoSM\ are quite similar. Then, stellar
masses are negligibly affected if \Mauto\ are adopted, instead of
aperture magnitudes. The impact of stellar population templates is
finally investigated for the BOSS datasample, comparing our
inferences with BOSS values from \cite{Maraston+13_BOSS}. These
BOSS masses are obtained fitting \cite{Maraston_Stromback11}
passive templates (with ages older than 3 Gyr) to the $\rm u, \,
g, \, r, \, i, \,z$ BOSS magnitudes. We have re-normalized
\cite{Maraston+13_BOSS} stellar masses to a Chabrier IMF, by
subtracting 0.05 dex, and converted to our cosmology by
subtracting 0.035 dex, too. On average, our stellar masses are
$\sim 0.08$ dex larger, this discrepancy is related to differences
in datasample and template prescription adopted. However, the
difference is also stronger at $z \lsim 0.5$, i.e. $\sim 0.12$
dex, and is inverted at $0.6 \lsim z \lsim 0.7$, where our masses
get smaller than BOSS estimates of $\sim 0.06$ dex. The trends of
\TtoSM\ stay qualitatively the same. Thus, the results of this
paper do not change if different stellar mass estimates are
adopted, and conclusions are unaffected.

\item[(iii)] {\it Stellar mass uncertainties.} The median
uncertainty of the Chabrier-IMF stellar masses determined in
\Sec\ref{subsec:st_masses} in our massive galaxy sample is $\sim
0.13$ dex. We studied the impact of these uncertainties on our
mass selection and all the trends discussed in the paper. We have
randomly perturbed each mass, assuming a Gaussian with as mean the
best-fitted mass and standard deviation the error of the
best-fitted mass. Thus, we have selected the most massive galaxies
with $\log \mst/\Msun > 11.2$, using the new perturbed masses. The
impact on the \TtoSM--z trend is negligible.

\item[(iv)] {\it Stellar \ML\ gradients.} We already shown how the gradients impact the results by comparing the g- and
K-band results for the SPIDER sample in
\Sec\ref{sec:DM_correlations}. However, in general, at masses
$\log \mst/\Msun > 11.2$ the colour and stellar \ML\ gradients are
weak, thus the correction would be small
(\citealt{Tortora+11MtoLgrad}). Later analysis would also take
into account possible gradients in the IMF
(\citealt{Martin-Navarro+15_IMF_variation}).

\item[(v)] {\it Mass profile slope.} In addition to uncertainties in stellar mass
estimates, the choice of the mass profile can be inappropriate for
lower-\sige\ (or \mst) galaxies and cause an excess of negative
\TtoSM\ values, mainly at high-$z$ (e.g.
\citealt{Sonnenfeld+13_SL2S_IV}). In particular, \TtoSM\ and the
slope of total mass density are tightly correlated, with shallower
density profiles corresponding to larger \TtoSM\
(\citealt{Humphrey_Buote10}; \citealt{Remus+13};
\citealt{Dutton_Treu14}; \citealt{Tortora+14_DMslope}). To
quantify the impact of a free-varying total mass density slope,
$\alpha$, we have adopted a power-law mass density $\rho \propto
r^{\alpha}$, with slope steeper and shallower than isothermal. In
\cite{Tortora+14_DMevol} we have shown that using two extreme
values for the density slope, i.e. $\alpha = -2.5$ and $-1.5$,
which bracket most of the results in the literature, then we find
that the average \TtoSM\ get smaller (larger) of $\sim 0.1$ ($\sim
0.05$) dex for $\alpha = -2.5$ ($=-1.5$). However, in the
realistic case with a varying slope with mass, smaller changes
would be expected (\citealt{Dutton_Treu14};
\citealt{Tortora+14_DMslope}). If $\alpha$ is constant with time,
then, at fixed mass, these corrections would shift all the results
of the same factor, leaving naturally unaffected the observed
trend with redshift. On the other hand, if we assume that the
slope is varying with redshift, then the only way to totally
remove any \TtoSM\ evolution is that low- and high- redshift
galaxies have systematically steeper (i.e. $\alpha <$ -2.5) and
shallower (i.e. $\alpha
>$ -1.5) mass
density profiles than isothermal, and consequently smaller and
larger \TtoSM. However, this is not expected since the most
massive galaxies are found to have only more shallow profiles than
isothermal at lower redshifts (e.g. \citealt{Tortora+14_DMslope};
\citealt{Poci+17_slope}) and at high redshifts, where gas and in
situ star formation dominate the galaxy evolution, the ETGs from
cosmological simulations have a total density slope very steep
($\alpha \sim -3$), with merging events driving the galaxy to a
nearly isothermal profile (\citealt{Remus+13,Remus+17}).
\cite{Xu+17_Illustris} from the Illustris simulation find a that
$\alpha$ is constant with redshift, which would leave the trend
unaffected. While, using strong lenses, \cite{Bolton+12_BELLS-II}
and \cite{Sonnenfeld+13_SL2S_IV} have shown that the total mass
density slope is steeper at lower redshift. In particular,
\cite{Sonnenfeld+13_SL2S_IV} find an average mass density slope of
$\alpha \sim -2$ at $z \sim 0.65$ and $\alpha \sim -2.15$ at $z
\sim 0.1$. Thus, at $z \sim 0.1$, \TtoSM\ would get smaller of
$\sim 0.03$ dex, leaving our trends almost unaffected. However, a
varying slope with redshift has to be further investigated.

\item[(vi)] {\it Rotation velocity. } We have also neglected rotation
correction, not accounted by the aperture velocity dispersion. If
rotation velocities would be included in the analysis \Mdyn\ would
get higher. But the impact would be negligible if compared with
typical uncertainties. In the massive and high velocity dispersion
galaxies analyzed in this paper, this change would amount to $\sim
1$ per cent (see \citealt{Tortora+09}).

\item[(vii)] {\it Orbital anisotropy.} The DM content of ETGs is
estimated under the assumption that their stellar orbits are
isotropic, which is incorrect at some level. Detailed dynamical
modeling analysis have focused on the ETG central regions, finding
anisotropies to be fairly mild in general, typically in the range
$-0.2 \le \beta \le +0.3$ (\citealt{Gerhard+01};
\citealt{Cappellari+07_SAURONX}), where $\beta \equiv 1 -
\sigma_\theta^2/\sigma_r^2$ quantifies the relative internal
dispersions in the tangential and radial directions. Mild central
anisotropy is also predicted from simulations of merger remnants
\citep{Dekel+05}. The typical radial anisotropies found in local
galaxies ($\beta \sim 0.2$), would decrease inferred dynamical
masses by $\lsim 5$ per cent (i.e. $\lsim 0.02$ dex), lowering the
\TtoSM. Recently, using a sample of massive ($\mst \gsim
10^{11}\Msun$) galaxies in the Illustris simulation,
\cite{Xu+17_Illustris} have predicted a gentle evolution in the
average central anisotropy of their sample, from $\beta \sim -0.1$
at $z \sim 0.65$ to $\beta \sim 0.1$ at $z\sim 0.1$. This means
that at $z\sim 0.1$ ($z \sim 0.65$) \TtoSM\ would be smaller
(larger) of $\lsim 3$ per cent ($\lsim 3$ per cent), flattening
our \TtoSM--z trend. However, the impact on our results is
negligible.

\item[(viii)] {\it Ellipticity.} We have checked the effect of the ellipticity in the
mass inferences and restricted the analysis to KiDS objects with
axis-ratio $q > 0.6$ ($86$ per cent of the sample), in order to
limit to the roundest galaxies. The overall results are
practically unchanged.

\item[(ix)] {\it S\'ersic index selection.} We have also checked if the results are affected if
we  select those objects with large S\'ersic indices (i.e.,
$r$-band $n > 2.5$), typical of most of the ETGs
(\citealt{Tortora+10CG}). This high--n sample consists of $\sim
94$ per cent of the original sample of massive galaxies. The
impact on the results is negligible, too.

\item[(x)] {\it Rest-frame structural parameters.} We have derived rest-frame structural
parameters by fitting a straight line $\log X = a + b \log
\lambda$ to the datapoints $(\lambda_{\rm l}, X_{\rm l})$ at the
three KiDS wavebands $g$, $r$ and $i$. To study the impact of this
choice, we have also replaced the straight line with the
polynomial function $\log X = a + b \log \lambda + c (\log
\lambda)^{2}$. The impact on our results is negligible, since the
difference between the derived S\'ersic index, effective radius
and \TtoSM\ using the two fitting functions are smaller than $3$
per cent.

\item[(xi)] We do not include in the mass model the black hole, but we have
calculated that its effect is, on average, negligible, with an
impact on \Mdyn\ of $\sim 2$ per cent (\citealt{Tortora+09}).
\end{itemize}

\section{Conclusions}\label{sec:conclusions}

We have analyzed the central DM content in a sample of massive
($\log \mst/\Msun > 11.2$) ETGs at redshift $\lsim 0.7$ from the
Kilo Degree Survey, a VST public survey
(\citealt{deJong+15_KiDS_paperI,deJong+17_KiDS_DR3}). Thanks to
the excellent seeing condition at Paranal, the small pixel scale
($0.2''$ per pixel) and the depth of the survey, KiDS provides us
with high quality images. We extract a sample of galaxies with
$\SN
> 50$, with masses and structural parameters from 156\sqd\ of the KiDS survey, and spectroscopic coverage from SDSS--DR7
and BOSS@SDSS datasamples. With spectroscopic redshifts,
structural parameters, stellar masses, and velocity dispersion for
a sample of 3778 massive ($\log\mst/\Msun > 11.2$) galaxies, we
can rely on the ideal datasample to perform the isotropic Jeans
dynamical analysis and study the central DM content, parameterized
through the total-to-stellar mass ratio, \TtoSM, (calculated at $r
= 1\, \rm \Re$; \citealt{Tortora+09,SPIDER-VI,Tortora+14_DMevol}).

We have shown that the local relationships between \TtoSM\ and
effective radius \Re, S\'ersic index n, velocity dispersion within
effective radius \sige, stellar mass \mst, central average stellar
density \rhostar\ and dynamical mass \Mdyn\ conserve their trends
at larger redshifts. These correlations are shown in
\Fig\ref{fig:DM}. We find that larger galaxies with steeper light
profiles (large n's) are more DM dominated. The effective radius
is the main driver of all the correlations. A similar steep trend
is found in terms of \Mdyn\ and \sigs. On the contrary, \TtoSM\
vary weakly in terms of \mst, while we find a steep
anti-correlation with central average stellar density. These
results confirm most of the previous literature at $z \sim 0$
(\citealt{Padmanabhan+04}; \citealt{Cappellari+06};
\citealt{HB09_curv}; \citealt{Tortora+09}; \citealt{NRT10};
\citealt{SPIDER-VI}), or at intermediate redshift
(\citealt{Tortora+10lensing}; \citealt{Auger+10_SLACSX};
\citealt{Barnabe+11}; \citealt{Tortora+14_DMevol}).

However, one of the most important result of this paper is that
the central DM is evolving from high$-$ to low$-z$, with less DM
at higher redshift. At fixed mass, we find that the galaxies in
the highest redshift bin have \TtoSM\ $\sim 0.2-0.4$ dex smaller
than local galaxies. Our findings are qualitatively consistent
with results in \cite{Beifiori+14} and \cite{Tortora+14_DMevol}.
We have also investigated the evolution of \Re, \sige\ and \TtoSM\
within various galaxy formation scenarios. We exclude the
"puffing-up" scenario from AGN feedback in \cite{Fan+08}, since
this kind of process produces very high \sige\ and \TtoSM\ at
higher--z, not observed. The predictions from galaxy merging
scenario from \cite{Hopkins+09_DELGN_IV} is consistent with the
observed size and mass growth with $z$ (\Fig\ref{fig:evolution}).

We find that the fraction of galaxies with negative DM is
increasing with redshift, but for the massive galaxies under
analysis a Chabrier or Salpeter IMF cannot be excluded. This
result is not in contrast with the recent findings in
\cite{Shetty_Cappellari14}, who find a Salpeter IMF in $z\sim
0.75$ massive galaxies. We have also shown (fixing the DM
fraction) how the trends of \TtoSM\ could be interpreted within
the context of a non-universal IMF, pointing to a varying IMF in
terms of the galaxy properties
(\citealt{TRN13_SPIDER_IMF,Tortora+14_DMslope,Tortora+14_MOND}).
If we assume a constant DM fraction with redshift, then the
observed change of \TtoSM\ is driven by an evolution of the IMF
(panels g and h of \Fig\ref{fig:evolution}). IMF is found to vary
from a Chabrier or Salpeter IMF at $z \sim 0.65$ to a
super-Salpeter or very steep IMF at low redshift. However, it is
difficult to explain this evolution within the galaxy formation
scenario, and in particular through galaxy mergers, since the
combination of a 'higher' (e.g., bottom-heavier) and a 'lower
mass' (e.g., bottom-lighter) IMF would produce a dilution in the
IMF shape, i.e. this process will not produce a more bottom-heavy
IMF, but an intermediate IMF slope. The mergers would decrease
\dimf\ or leave it almost constant with redshift, as predicted by
the galaxy merging toy-models in \cite{Sonnenfeld+17_IMF},
contrasting the evolution seen in \Fig\ref{fig:evolution}, and
suggesting that \fdm\ need to evolve with cosmic time. However, a
detailed analysis of the joint IMF and DM fraction evolution
across the time is beyond the scope of this paper. We will address
this issue in future, assuming more complex mass modelling
(\citealt{TRN13_SPIDER_IMF,
Tortora+14_MOND,Tortora+14_DMslope,TLBN16_IMF_dwarfs}).

Finally, to have a more conclusive answer about the physical
processes leading the evolution of these massive galaxies, we have
studied how \Re--\mst\ and \TtoSM--\mst\ correlations change with
redshift. And we have used toy-models to investigate how this
evolution can be explained by the mass accretion from minor- and
major-mergings (\citealt{Boylan-Kolchin+05}; \citealt{Hilz+13};
\citealt{Tortora+14_DMevol}). We find that our results point to
minor mergers as main driver of the  mass accretion, since major
mergers would produce a too strong mass accretion, which is not
observed (\Fig\ref{fig:evol_toy_models}, the size evolution is
parallel to the \Re--\mst\ relation for galaxies at the same
redshift). To consistently reproduce the evolution in both the
size and \TtoSM, then, we have demonstrated that the accretion of
total DM mass would be weaker than the stellar mass one. This can
be achieved if the main galaxy is merging with lower-mass galaxies
with a smaller \TtoSM. This result is consistent with independent
studies which find that \TtoSM\ is getting smaller at lower
masses, down to the threshold mass of $\sim 10^{10.5}\, \Msun$,
translating to a star formation efficiency which is increasing
from large and massive galaxies to intermediate-mass galaxies
(\citealt{Benson+00}; \citealt{MH02}; \citealt{Napolitano+05};
\citealt{Mandelbaum+06}; \citealt{vdB+07}; \citealt{CW09};
\citealt{Moster+10}).

In a future work we will perform a more systematic study of the
IMF, to investigate if it can be considered universal with cosmic
time, or if it is changing as a function of the redshift. Then, we
plan to have 10 times more galaxies at the end of the KiDS survey,
when all the 1500\sqd\ will have been observed.


\section*{Acknowledgments}
We thank the anonymous referee for his/her comments. CT and LVEK
are supported through an NWO-VICI grant (project number
639.043.308). KK acknowledges support by the Alexander von
Humboldt Foundation. GVK acknowledges financial support from the
Netherlands Research School for Astronomy (NOVA) and Target.
Target is supported by Samenwerkingsverband Noord Nederland,
European fund for regional development, Dutch Ministry of economic
affairs, Pieken in de Delta, Provinces of Groningen and Drenthe.
Based on data products from observations made with ESO Telescopes
at the La Silla Paranal Observatory under programme IDs
177.A-3016, 177.A-3017 and 177.A-3018, and on data products
produced by Target/OmegaCEN, INAF-OACN, INAF-OAPD and the KiDS
production team, on behalf of the KiDS consortium. OmegaCEN and
the KiDS production team acknowledge support by NOVA and NWO-M
grants. Members of INAF-OAPD and INAF-OACN also acknowledge the
support from the Department of Physics \& Astronomy of the
University of Padova, and of the Department of Physics of Univ.
Federico II (Naples).


\bibliographystyle{mn2e}   



\begin{thebibliography}{150}
\expandafter\ifx\csname
natexlab\endcsname\relax\def\natexlab#1{#1}\fi

\bibitem[{{Abazajian} {et~al}\mbox{.}(2003){Abazajian}, {Adelman-McCarthy},
  {Ag{\"u}eros}, {Allam}, {Anderson}, {Annis}, {Bahcall}, {Baldry}, {Bastian},
  {Berlind}, {Bernardi}, {Blanton}, {Blythe}, {Bochanski}, {Boroski},
  {Brewington}, {Briggs}, {Brinkmann}, {Brunner}, {Budav{\'a}ri}, {Carey},
  {Carr}, {Castander}, {Chiu}, {Collinge}, {Connolly}, {Covey}, {Csabai},
  {Dalcanton}, {Dodelson}, {Doi}, {Dong}, {Eisenstein}, {Evans}, {Fan},
  {Feldman}, {Finkbeiner}, {Friedman}, {Frieman}, {Fukugita}, {Gal},
  {Gillespie}, {Glazebrook}, {Gonzalez}, {Gray}, {Grebel}, {Grodnicki}, {Gunn},
  {Gurbani}, {Hall}, {Hao}, {Harbeck}, {Harris}, {Harris}, {Harvanek},
  {Hawley}, {Heckman}, {Helmboldt}, {Hendry}, {Hennessy}, {Hindsley}, {Hogg},
  {Holmgren}, {Holtzman}, {Homer}, {Hui}, {Ichikawa}, {Ichikawa}, {Inkmann},
  {Ivezi{\'c}}, {Jester}, {Johnston}, {Jordan}, {Jordan}, {Jorgensen},
  {Juri{\'c}}, {Kauffmann}, {Kent}, {Kleinman}, {Knapp}, {Kniazev}, {Kron},
  {Krzesi{\'n}ski}, {Kunszt}, {Kuropatkin}, {Lamb}, {Lampeitl}, {Laubscher},
  {Lee}, {Leger}, {Li}, {Lidz}, {Lin}, {Loh}, {Long}, {Loveday}, {Lupton},
  {Malik}, {Margon}, {McGehee}, {McKay}, {Meiksin}, {Miknaitis}, {Moorthy},
  {Munn}, {Murphy}, {Nakajima}, {Narayanan}, {Nash}, {Neilsen}, {Newberg},
  {Newman}, {Nichol}, {Nicinski}, {Nieto-Santisteban}, {Nitta}, {Odenkirchen},
  {Okamura}, {Ostriker}, {Owen}, {Padmanabhan}, {Peoples}, {Pier}, {Pindor},
  {Pope}, {Quinn}, {Rafikov}, {Raymond}, {Richards}, {Richmond}, {Rix},
  {Rockosi}, {Schaye}, {Schlegel}, {Schneider}, {Schroeder}, {Scranton},
  {Sekiguchi}, {Seljak}, {Sergey}, {Sesar}, {Sheldon}, {Shimasaku}, {Siegmund},
  {Silvestri}, {Sinisgalli}, {Sirko}, {Smith}, {Smol{\v c}i{\'c}}, {Snedden},
  {Stebbins}, {Steinhardt}, {Stinson}, {Stoughton}, {Strateva}, {Strauss},
  {SubbaRao}, {Szalay}, {Szapudi}, {Szkody}, {Tasca}, {Tegmark}, {Thakar},
  {Tremonti}, {Tucker}, {Uomoto}, {Vanden Berk}, {Vandenberg}, {Vogeley},
  {Voges}, {Vogt}, {Walkowicz}, {Weinberg}, {West}, {White}, {Wilhite},
  {Willman}, {Xu}, {Yanny}, {Yarger}, {Yasuda}, {Yip}, {Yocum}, {York},
  {Zakamska}, {Zehavi}, {Zheng}, {Zibetti}, \& {Zucker}}]{SDSS_DR1}
{Abazajian} K. {et~al.}, 2003, \aj, 126, 2081

\bibitem[{{Abazajian} {et~al}\mbox{.}(2009){Abazajian}, {Adelman-McCarthy},
  {Ag{\"u}eros}, {Allam}, {Allende Prieto}, {An}, {Anderson}, {Anderson},
  {Annis}, {Bahcall}, \& et~al.}]{SDSS_DR7_Abazajian}
{Abazajian} K.~N. {et~al.}, 2009, \apjs, 182, 543

\bibitem[{{Adelman-McCarthy} {et~al}\mbox{.}(2008){Adelman-McCarthy},
  {Ag{\"u}eros}, {Allam}, {Allende Prieto}, {Anderson}, {Anderson}, {Annis},
  {Bahcall}, {Bailer-Jones}, {Baldry}, {Barentine}, {Bassett}, {Becker},
  {Beers}, {Bell}, {Berlind}, {Bernardi}, {Blanton}, {Bochanski}, {Boroski},
  {Brinchmann}, {Brinkmann}, {Brunner}, {Budav{\'a}ri}, {Carliles}, {Carr},
  {Castander}, {Cinabro}, {Cool}, {Covey}, {Csabai}, {Cunha}, {Davenport},
  {Dilday}, {Doi}, {Eisenstein}, {Evans}, {Fan}, {Finkbeiner}, {Friedman},
  {Frieman}, {Fukugita}, {G{\"a}nsicke}, {Gates}, {Gillespie}, {Glazebrook},
  {Gray}, {Grebel}, {Gunn}, {Gurbani}, {Hall}, {Harding}, {Harvanek}, {Hawley},
  {Hayes}, {Heckman}, {Hendry}, {Hindsley}, {Hirata}, {Hogan}, {Hogg}, {Hyde},
  {Ichikawa}, {Ivezi{\'c}}, {Jester}, {Johnson}, {Jorgensen}, {Juri{\'c}},
  {Kent}, {Kessler}, {Kleinman}, {Knapp}, {Kron}, {Krzesinski}, {Kuropatkin},
  {Lamb}, {Lampeitl}, {Lebedeva}, {Lee}, {Leger}, {L{\'e}pine}, {Lima}, {Lin},
  {Long}, {Loomis}, {Loveday}, {Lupton}, {Malanushenko}, {Malanushenko},
  {Mandelbaum}, {Margon}, {Marriner}, {Mart{\'{\i}}nez-Delgado}, {Matsubara},
  {McGehee}, {McKay}, {Meiksin}, {Morrison}, {Munn}, {Nakajima}, {Neilsen},
  {Newberg}, {Nichol}, {Nicinski}, {Nieto-Santisteban}, {Nitta}, {Okamura},
  {Owen}, {Oyaizu}, {Padmanabhan}, {Pan}, {Park}, {Peoples}, {Pier}, {Pope},
  {Purger}, {Raddick}, {Re Fiorentin}, {Richards}, {Richmond}, {Riess}, {Rix},
  {Rockosi}, {Sako}, {Schlegel}, {Schneider}, {Schreiber}, {Schwope}, {Seljak},
  {Sesar}, {Sheldon}, {Shimasaku}, {Sivarani}, {Smith}, {Snedden}, {Steinmetz},
  {Strauss}, {SubbaRao}, {Suto}, {Szalay}, {Szapudi}, {Szkody}, {Tegmark},
  {Thakar}, {Tremonti}, {Tucker}, {Uomoto}, {Vanden Berk}, {Vandenberg},
  {Vidrih}, {Vogeley}, {Voges}, {Vogt}, {Wadadekar}, {Weinberg}, {West},
  {White}, {Wilhite}, {Yanny}, {Yocum}, {York}, {Zehavi}, \&
  {Zucker}}]{SDSS_DR6}
{Adelman-McCarthy} J.~K. {et~al.}, 2008, \apjs, 175, 297

\bibitem[{{Ahn} {et~al}\mbox{.}(2014){Ahn}, {Alexandroff}, {Allende Prieto},
  {Anders}, {Anderson}, {Anderton}, {Andrews}, {Aubourg}, {Bailey}, {Bastien},
  \& et~al.}]{Ahn+14_SDSS_DR10}
{Ahn} C.~P. {et~al.}, 2014, \apjs, 211, 17

\bibitem[{{Alabi} {et~al}\mbox{.}(2016){Alabi}, {Forbes}, {Romanowsky},
  {Brodie}, {Strader}, {Janz}, {Pota}, {Pastorello}, {Usher}, {Spitler},
  {Foster}, {Jennings}, {Villaume}, \& {Kartha}}]{Alabi+16}
{Alabi} A.~B. {et~al.}, 2016, \mnras, 460, 3838

\bibitem[{{Arnouts} {et~al}\mbox{.}(1999){Arnouts}, {Cristiani}, {Moscardini},
  {Matarrese}, {Lucchin}, {Fontana}, \& {Giallongo}}]{Arnouts+99}
{Arnouts} S., {Cristiani} S., {Moscardini} L., {Matarrese} S.,
{Lucchin} F.,
  {Fontana} A., {Giallongo} E., 1999, \mnras, 310, 540

\bibitem[{{Auger} {et~al}\mbox{.}(2009){Auger}, {Treu}, {Bolton}, {Gavazzi},
  {Koopmans}, {Marshall}, {Bundy}, \& {Moustakas}}]{Auger+09_SLACSIX}
{Auger} M.~W., {Treu} T., {Bolton} A.~S., {Gavazzi} R., {Koopmans}
L.~V.~E.,
  {Marshall} P.~J., {Bundy} K., {Moustakas} L.~A., 2009, \apj, 705, 1099

\bibitem[{{Auger} {et~al}\mbox{.}(2010){Auger}, {Treu}, {Bolton}, {Gavazzi},
  {Koopmans}, {Marshall}, {Moustakas}, \& {Burles}}]{Auger+10_SLACSX}
{Auger} M.~W., {Treu} T., {Bolton} A.~S., {Gavazzi} R., {Koopmans}
L.~V.~E.,
  {Marshall} P.~J., {Moustakas} L.~A., {Burles} S., 2010, \apj, 724, 511

\bibitem[{{Barnab{\`e}} {et~al}\mbox{.}(2011){Barnab{\`e}}, {Czoske},
  {Koopmans}, {Treu}, \& {Bolton}}]{Barnabe+11}
{Barnab{\`e}} M., {Czoske} O., {Koopmans} L.~V.~E., {Treu} T.,
{Bolton} A.~S.,
  2011, \mnras, 415, 2215

\bibitem[{{Barnab{\`e}} {et~al}\mbox{.}(2013){Barnab{\`e}}, {Spiniello},
  {Koopmans}, {Trager}, {Czoske}, \& {Treu}}]{Barnabe+13}
{Barnab{\`e}} M., {Spiniello} C., {Koopmans} L.~V.~E., {Trager}
S.~C., {Czoske}
  O., {Treu} T., 2013, \mnras, 436, 253

\bibitem[{{Bastian} {et~al}\mbox{.}(2010){Bastian}, {Covey}, \&
  {Meyer}}]{Bastian+10}
{Bastian} N., {Covey} K.~R., {Meyer} M.~R., 2010, \araa, 48, 339

\bibitem[{{Beifiori} {et~al}\mbox{.}(2014){Beifiori}, {Thomas}, {Maraston},
  {Steele}, {Masters}, {Pforr}, {Saglia}, {Bender}, {Tojeiro}, {Chen},
  {Bolton}, {Brownstein}, {Johansson}, {Leauthaud}, {Nichol}, {Schneider},
  {Senger}, {Skibba}, {Wake}, {Pan}, {Snedden}, {Bizyaev}, {Brewington},
  {Malanushenko}, {Malanushenko}, {Oravetz}, {Simmons}, {Shelden}, \&
  {Ebelke}}]{Beifiori+14}
{Beifiori} A. {et~al.}, 2014, \apj, 789, 92

\bibitem[{{Benson} {et~al}\mbox{.}(2000){Benson}, {Cole}, {Frenk}, {Baugh}, \&
  {Lacey}}]{Benson+00}
{Benson} A.~J., {Cole} S., {Frenk} C.~S., {Baugh} C.~M., {Lacey}
C.~G., 2000,
  \mnras, 311, 793

\bibitem[{{Blumenthal} {et~al}\mbox{.}(1984){Blumenthal}, {Faber}, {Primack},
  \& {Rees}}]{Blumenthal+84_nature}
{Blumenthal} G.~R., {Faber} S.~M., {Primack} J.~R., {Rees} M.~J.,
1984, \nat,
  311, 517

\bibitem[{{Bolton} {et~al}\mbox{.}(2012){Bolton}, {Brownstein}, {Kochanek},
  {Shu}, {Schlegel}, {Eisenstein}, {Wake}, {Connolly}, {Maraston}, {Arneson},
  \& {Weaver}}]{Bolton+12_BELLS-II}
{Bolton} A.~S. {et~al.}, 2012, \apj, 757, 82

\bibitem[{{Bolton} {et~al}\mbox{.}(2008){Bolton}, {Burles}, {Koopmans}, {Treu},
  {Gavazzi}, {Moustakas}, {Wayth}, \& {Schlegel}}]{Bolton+08_SLACSV}
{Bolton} A.~S., {Burles} S., {Koopmans} L.~V.~E., {Treu} T.,
{Gavazzi} R.,
  {Moustakas} L.~A., {Wayth} R., {Schlegel} D.~J., 2008, \apj, 682, 964

\bibitem[{{Bolton} {et~al}\mbox{.}(2006){Bolton}, {Burles}, {Koopmans}, {Treu},
  \& {Moustakas}}]{Bolton+06_SLACSI}
{Bolton} A.~S., {Burles} S., {Koopmans} L.~V.~E., {Treu} T.,
{Moustakas} L.~A.,
  2006, \apj, 638, 703

\bibitem[{{Boylan-Kolchin} {et~al}\mbox{.}(2005){Boylan-Kolchin}, {Ma}, \&
  {Quataert}}]{Boylan-Kolchin+05}
{Boylan-Kolchin} M., {Ma} C.-P., {Quataert} E., 2005, \mnras, 362,
184

\bibitem[{{Bruzual} \& {Charlot}(2003)}]{BC03}
{Bruzual} G., {Charlot} S., 2003, \mnras, 344, 1000

\bibitem[{{Buitrago} {et~al}\mbox{.}(2008){Buitrago}, {Trujillo}, {Conselice},
  {Bouwens}, {Dickinson}, \& {Yan}}]{Buitrago+08}
{Buitrago} F., {Trujillo} I., {Conselice} C.~J., {Bouwens} R.~J.,
{Dickinson}
  M., {Yan} H., 2008, \apjl, 687, L61

\bibitem[{{Bullock} {et~al}\mbox{.}(2001){Bullock}, {Kolatt}, {Sigad},
  {Somerville}, {Kravtsov}, {Klypin}, {Primack}, \& {Dekel}}]{Bullock+01}
{Bullock} J.~S., {Kolatt} T.~S., {Sigad} Y., {Somerville} R.~S.,
{Kravtsov}
  A.~V., {Klypin} A.~A., {Primack} J.~R., {Dekel} A., 2001, \mnras, 321, 559

\bibitem[{{Cappellari} {et~al}\mbox{.}(2006){Cappellari}, {Bacon}, {Bureau},
  {Damen}, {Davies}, {de Zeeuw}, {Emsellem}, {Falc{\'o}n-Barroso},
  {Krajnovi{\'c}}, {Kuntschner}, {McDermid}, {Peletier}, {Sarzi}, {van den
  Bosch}, \& {van de Ven}}]{Cappellari+06}
{Cappellari} M. {et~al.}, 2006, \mnras, 366, 1126

\bibitem[{{Cappellari} \& {Emsellem}(2004)}]{Cappellari_Emsellem14}
{Cappellari} M., {Emsellem} E., 2004, \pasp, 116, 138

\bibitem[{{Cappellari} {et~al}\mbox{.}(2007){Cappellari}, {Emsellem}, {Bacon},
  {Bureau}, {Davies}, {de Zeeuw}, {Falc{\'o}n-Barroso}, {Krajnovi{\'c}},
  {Kuntschner}, {McDermid}, {Peletier}, {Sarzi}, {van den Bosch}, \& {van de
  Ven}}]{Cappellari+07_SAURONX}
{Cappellari} M. {et~al.}, 2007, \mnras, 379, 418

\bibitem[{{Cappellari} {et~al}\mbox{.}(2012){Cappellari}, {McDermid},
  {Alatalo}, {Blitz}, {Bois}, {Bournaud}, {Bureau}, {Crocker}, {Davies},
  {Davis}, {de Zeeuw}, {Duc}, {Emsellem}, {Khochfar}, {Krajnovi{\'c}},
  {Kuntschner}, {Lablanche}, {Morganti}, {Naab}, {Oosterloo}, {Sarzi}, {Scott},
  {Serra}, {Weijmans}, \& {Young}}]{Cappellari+12}
{Cappellari} M. {et~al.}, 2012, \nat, 484, 485

\bibitem[{{Cappellari} {et~al}\mbox{.}(2013){Cappellari}, {McDermid},
  {Alatalo}, {Blitz}, {Bois}, {Bournaud}, {Bureau}, {Crocker}, {Davies},
  {Davis}, {de Zeeuw}, {Duc}, {Emsellem}, {Khochfar}, {Krajnovi{\'c}},
  {Kuntschner}, {Morganti}, {Naab}, {Oosterloo}, {Sarzi}, {Scott}, {Serra},
  {Weijmans}, \& {Young}}]{Cappellari+13_ATLAS3D_XX}
{Cappellari} M. {et~al.}, 2013, \mnras, 432, 1862

\bibitem[{{Cardone} {et~al}\mbox{.}(2011){Cardone}, {Del Popolo}, {Tortora}, \&
  {Napolitano}}]{Cardone+11SIM}
{Cardone} V.~F., {Del Popolo} A., {Tortora} C., {Napolitano}
N.~R., 2011,
  \mnras, 416, 1822

\bibitem[{{Cardone} \& {Tortora}(2010)}]{CT10}
{Cardone} V.~F., {Tortora} C., 2010, \mnras, 409, 1570

\bibitem[{{Cardone} {et~al}\mbox{.}(2009){Cardone}, {Tortora}, {Molinaro}, \&
  {Salzano}}]{Cardone+09}
{Cardone} V.~F., {Tortora} C., {Molinaro} R., {Salzano} V., 2009,
\aap, 504,
  769

\bibitem[{{Cenarro} \& {Trujillo}(2009)}]{Cenarro_Trujillo09}
{Cenarro} A.~J., {Trujillo} I., 2009, \apjl, 696, L43

\bibitem[{{Chabrier}(2001)}]{Chabrier01}
{Chabrier} G., 2001, \apj, 554, 1274

\bibitem[{{Chae} {et~al}\mbox{.}(2014){Chae}, {Bernardi}, \&
  {Kravtsov}}]{Chae+14}
{Chae} K.-H., {Bernardi} M., {Kravtsov} A.~V., 2014, \mnras, 437,
3670

\bibitem[{{Conroy} \& {van Dokkum}(2012)}]{Conroy_vanDokkum12b}
{Conroy} C., {van Dokkum} P.~G., 2012, \apj, 760, 71

\bibitem[{{Conroy} \& {Wechsler}(2009)}]{CW09}
{Conroy} C., {Wechsler} R.~H., 2009, \apj, 696, 620

\bibitem[{{Conselice}(2014)}]{Conselice14_review}
{Conselice} C.~J., 2014, \araa, 52, 291

\bibitem[{{Corsini} {et~al}\mbox{.}(2017){Corsini}, {Wegner}, {Thomas},
  {Saglia}, \& {Bender}}]{Corsini+17}
{Corsini} E.~M., {Wegner} G.~A., {Thomas} J., {Saglia} R.~P.,
{Bender} R.,
  2017, \mnras, 466, 974

\bibitem[{{Daddi} {et~al}\mbox{.}(2005){Daddi}, {Renzini}, {Pirzkal},
  {Cimatti}, {Malhotra}, {Stiavelli}, {Xu}, {Pasquali}, {Rhoads}, {Brusa}, {di
  Serego Alighieri}, {Ferguson}, {Koekemoer}, {Moustakas}, {Panagia}, \&
  {Windhorst}}]{Daddi+05}
{Daddi} E. {et~al.}, 2005, \apj, 626, 680

\bibitem[{{de Jong} {et~al}\mbox{.}(2017){de Jong}, {Kleijn}, {Erben},
  {Hildebrandt}, {Kuijken}, {Sikkema}, {Brescia}, {Bilicki}, {Napolitano},
  {Amaro}, {Begeman}, {Boxhoorn}, {Buddelmeijer}, {Cavuoti}, {Getman}, {Grado},
  {Helmich}, {Huang}, {Irisarri}, {La Barbera}, {Longo}, {McFarland},
  {Nakajima}, {Paolillo}, {Puddu}, {Radovich}, {Rifatto}, {Tortora},
  {Valentijn}, {Vellucci}, {Vriend}, {Amon}, {Blake}, {Choi}, {Conti}, {Gwyn},
  {Herbonnet}, {Heymans}, {Hoekstra}, {Klaes}, {Merten}, {Miller}, {Schneider},
  \& {Viola}}]{deJong+17_KiDS_DR3}
{de Jong} J.~T.~A. {et~al.}, 2017, \aap, 604, A134

\bibitem[{{de Jong} {et~al}\mbox{.}(2015){de Jong}, {Verdoes Kleijn},
  {Boxhoorn}, {Buddelmeijer}, {Capaccioli}, {Getman}, {Grado}, {Helmich},
  {Huang}, {Irisarri}, {Kuijken}, {La Barbera}, {McFarland}, {Napolitano},
  {Radovich}, {Sikkema}, {Valentijn}, {Begeman}, {Brescia}, {Cavuoti}, {Choi},
  {Cordes}, {Covone}, {Dall'Ora}, {Hildebrandt}, {Longo}, {Nakajima},
  {Paolillo}, {Puddu}, {Rifatto}, {Tortora}, {van Uitert}, {Buddendiek},
  {Harnois-D{\'e}raps}, {Erben}, {Eriksen}, {Heymans}, {Hoekstra}, {Joachimi},
  {Kitching}, {Klaes}, {Koopmans}, {K{\"o}hlinger}, {Roy}, {Sif{\'o}n},
  {Schneider}, {Sutherland}, {Viola}, \& {Vriend}}]{deJong+15_KiDS_paperI}
{de Jong} J.~T.~A. {et~al.}, 2015, \aap, 582, A62

\bibitem[{{de Vaucouleurs}(1948)}]{deVauc48}
{de Vaucouleurs} G., 1948, Annales d'Astrophysique, 11, 247

\bibitem[{{Dekel} {et~al}\mbox{.}(2005){Dekel}, {Stoehr}, {Mamon}, {Cox},
  {Novak}, \& {Primack}}]{Dekel+05}
{Dekel} A., {Stoehr} F., {Mamon} G.~A., {Cox} T.~J., {Novak}
G.~S., {Primack}
  J.~R., 2005, \nat, 437, 707

\bibitem[{{Dutton} {et~al}\mbox{.}(2013){Dutton}, {Macci{\`o}}, {Mendel}, \&
  {Simard}}]{Dutton+13}
{Dutton} A.~A., {Macci{\`o}} A.~V., {Mendel} J.~T., {Simard} L.,
2013, \mnras,
  432, 2496

\bibitem[{{Dutton} \& {Treu}(2014)}]{Dutton_Treu14}
{Dutton} A.~A., {Treu} T., 2014, \mnras, 438, 3594

\bibitem[{{Eisenstein} {et~al}\mbox{.}(2011){Eisenstein}, {Weinberg}, {Agol},
  {Aihara}, {Allende Prieto}, {Anderson}, {Arns}, {Aubourg}, {Bailey},
  {Balbinot}, \& et~al.}]{Eisenstein+11_SDSSIII}
{Eisenstein} D.~J. {et~al.}, 2011, \aj, 142, 72

\bibitem[{{Fan} {et~al}\mbox{.}(2008){Fan}, {Lapi}, {De Zotti}, \&
  {Danese}}]{Fan+08}
{Fan} L., {Lapi} A., {De Zotti} G., {Danese} L., 2008, \apjl, 689,
L101

\bibitem[{{Faure} {et~al}\mbox{.}(2011){Faure}, {Anguita}, {Alloin}, {Bundy},
  {Finoguenov}, {Leauthaud}, {Knobel}, {Kneib}, {Jullo}, {Ilbert}, {Koekemoer},
  {Capak}, {Scoville}, \& {Tasca}}]{Faure+11}
{Faure} C. {et~al.}, 2011, \aap, 529, A72

\bibitem[{{Ferreras} {et~al}\mbox{.}(2013){Ferreras}, {La Barbera}, {de la
  Rosa}, {Vazdekis}, {de Carvalho}, {Falc{\'o}n-Barroso}, \&
  {Ricciardelli}}]{Ferreras+13}
{Ferreras} I., {La Barbera} F., {de la Rosa} I.~G., {Vazdekis} A.,
{de
  Carvalho} R.~R., {Falc{\'o}n-Barroso} J., {Ricciardelli} E., 2013, \mnras,
  429, L15

\bibitem[{{Gallazzi} {et~al}\mbox{.}(2005){Gallazzi}, {Charlot}, {Brinchmann},
  {White}, \& {Tremonti}}]{Gallazzi+05}
{Gallazzi} A., {Charlot} S., {Brinchmann} J., {White} S.~D.~M.,
{Tremonti}
  C.~A., 2005, \mnras, 362, 41

\bibitem[{{Gavazzi} {et~al}\mbox{.}(2007){Gavazzi}, {Treu}, {Rhodes},
  {Koopmans}, {Bolton}, {Burles}, {Massey}, \&
  {Moustakas}}]{Gavazzi+07_SLACSIV}
{Gavazzi} R., {Treu} T., {Rhodes} J.~D., {Koopmans} L.~V.~E.,
{Bolton} A.~S.,
  {Burles} S., {Massey} R.~J., {Moustakas} L.~A., 2007, \apj, 667, 176

\bibitem[{{Gerhard} {et~al}\mbox{.}(2001){Gerhard}, {Kronawitter}, {Saglia}, \&
  {Bender}}]{Gerhard+01}
{Gerhard} O., {Kronawitter} A., {Saglia} R.~P., {Bender} R., 2001,
\aj, 121,
  1936

\bibitem[{{Gnedin} {et~al}\mbox{.}(2004){Gnedin}, {Kravtsov}, {Klypin}, \&
  {Nagai}}]{Gnedin+04}
{Gnedin} O.~Y., {Kravtsov} A.~V., {Klypin} A.~A., {Nagai} D.,
2004, \apj, 616,
  16

\bibitem[{{Gnedin} {et~al}\mbox{.}(2007){Gnedin}, {Weinberg}, {Pizagno},
  {Prada}, \& {Rix}}]{Gnedin+07}
{Gnedin} O.~Y., {Weinberg} D.~H., {Pizagno} J., {Prada} F., {Rix}
H.-W., 2007,
  \apj, 671, 1115

\bibitem[{{Goudfrooij} \& {Kruijssen}(2013)}]{Goudfrooij_Kruijssen13}
{Goudfrooij} P., {Kruijssen} J.~M.~D., 2013, \apj, 762, 107

\bibitem[{{Goudfrooij} \& {Kruijssen}(2014)}]{Goudfrooij_Kruijssen14}
{Goudfrooij} P., {Kruijssen} J.~M.~D., 2014, \apj, 780, 43

\bibitem[{{Graves} {et~al}\mbox{.}(2009){Graves}, {Faber}, \&
  {Schiavon}}]{Graves+09}
{Graves} G.~J., {Faber} S.~M., {Schiavon} R.~P., 2009, \apj, 698,
1590

\bibitem[{{Grillo}(2010)}]{Grillo10}
{Grillo} C., 2010, \apj, 722, 779

\bibitem[{{Grillo} \& {Gobat}(2010)}]{Grillo_Cobat10}
{Grillo} C., {Gobat} R., 2010, \mnras, 402, L67

\bibitem[{{Grillo} {et~al}\mbox{.}(2009){Grillo}, {Gobat}, {Lombardi}, \&
  {Rosati}}]{Grillo+09}
{Grillo} C., {Gobat} R., {Lombardi} M., {Rosati} P., 2009, \aap,
501, 461

\bibitem[{{Hilz} {et~al}\mbox{.}(2013){Hilz}, {Naab}, \& {Ostriker}}]{Hilz+13}
{Hilz} M., {Naab} T., {Ostriker} J.~P., 2013, \mnras, 429, 2924

\bibitem[{{Hopkins} {et~al}\mbox{.}(2009){Hopkins}, {Hernquist}, {Cox},
  {Keres}, \& {Wuyts}}]{Hopkins+09_DELGN_IV}
{Hopkins} P.~F., {Hernquist} L., {Cox} T.~J., {Keres} D., {Wuyts}
S., 2009,
  \apj, 691, 1424

\bibitem[{{Humphrey} \& {Buote}(2010)}]{Humphrey_Buote10}
{Humphrey} P.~J., {Buote} D.~A., 2010, \mnras, 403, 2143

\bibitem[{{Hyde} \& {Bernardi}(2009{\natexlab{a}})}]{HB09_curv}
{Hyde} J.~B., {Bernardi} M., 2009{\natexlab{a}}, \mnras, 394, 1978

\bibitem[{{Hyde} \& {Bernardi}(2009{\natexlab{b}})}]{HB09_FP}
{Hyde} J.~B., {Bernardi} M., 2009{\natexlab{b}}, \mnras, 396, 1171

\bibitem[{{Ilbert} {et~al}\mbox{.}(2006){Ilbert}, {Arnouts}, {McCracken},
  {Bolzonella}, {Bertin}, {Le F{\`e}vre}, {Mellier}, {Zamorani}, {Pell{\`o}},
  {Iovino}, {Tresse}, {Le Brun}, {Bottini}, {Garilli}, {Maccagni}, {Picat},
  {Scaramella}, {Scodeggio}, {Vettolani}, {Zanichelli}, {Adami}, {Bardelli},
  {Cappi}, {Charlot}, {Ciliegi}, {Contini}, {Cucciati}, {Foucaud}, {Franzetti},
  {Gavignaud}, {Guzzo}, {Marano}, {Marinoni}, {Mazure}, {Meneux}, {Merighi},
  {Paltani}, {Pollo}, {Pozzetti}, {Radovich}, {Zucca}, {Bondi}, {Bongiorno},
  {Busarello}, {de La Torre}, {Gregorini}, {Lamareille}, {Mathez}, {Merluzzi},
  {Ripepi}, {Rizzo}, \& {Vergani}}]{Ilbert+06}
{Ilbert} O. {et~al.}, 2006, \aap, 457, 841

\bibitem[{{Jorgensen} {et~al}\mbox{.}(1995){Jorgensen}, {Franx}, \&
  {Kjaergaard}}]{JFK95}
{Jorgensen} I., {Franx} M., {Kjaergaard} P., 1995, \mnras, 273,
1097

\bibitem[{{Khochfar} \& {Silk}(2006)}]{Khochfar_Silk06}
{Khochfar} S., {Silk} J., 2006, \apjl, 648, L21

\bibitem[{{Kochanek}(1991)}]{Kochanek91}
{Kochanek} C.~S., 1991, \apj, 373, 354

\bibitem[{{Komatsu} {et~al}\mbox{.}(2011){Komatsu}, {Smith}, {Dunkley},
  {Bennett}, {Gold}, {Hinshaw}, {Jarosik}, {Larson}, {Nolta}, {Page},
  {Spergel}, {Halpern}, {Hill}, {Kogut}, {Limon}, {Meyer}, {Odegard}, {Tucker},
  {Weiland}, {Wollack}, \& {Wright}}]{Komatsu+11_WMAP7}
{Komatsu} E. {et~al.}, 2011, \apjs, 192, 18

\bibitem[{{Koopmans} {et~al}\mbox{.}(2006){Koopmans}, {Treu}, {Bolton},
  {Burles}, \& {Moustakas}}]{Koopmans+06_SLACSIII}
{Koopmans} L.~V.~E., {Treu} T., {Bolton} A.~S., {Burles} S.,
{Moustakas} L.~A.,
  2006, \apj, 649, 599

\bibitem[{{La Barbera} \& {de Carvalho}(2009)}]{LaBarbera_deCarvalho09}
{La Barbera} F., {de Carvalho} R.~R., 2009, \apjl, 699, L76

\bibitem[{{La Barbera} {et~al}\mbox{.}(2010){La Barbera}, {de Carvalho}, {de La
  Rosa}, {Lopes}, {Kohl-Moreira}, \& {Capelato}}]{SPIDER-I}
{La Barbera} F., {de Carvalho} R.~R., {de La Rosa} I.~G., {Lopes}
P.~A.~A.,
  {Kohl-Moreira} J.~L., {Capelato} H.~V., 2010, \mnras, 408, 1313

\bibitem[{{La Barbera} {et~al}\mbox{.}(2008){La Barbera}, {de Carvalho},
  {Kohl-Moreira}, {Gal}, {Soares-Santos}, {Capaccioli}, {Santos}, \&
  {Sant'anna}}]{LaBarbera_08_2DPHOT}
{La Barbera} F., {de Carvalho} R.~R., {Kohl-Moreira} J.~L., {Gal}
R.~R.,
  {Soares-Santos} M., {Capaccioli} M., {Santos} R., {Sant'anna} N., 2008,
  \pasp, 120, 681

\bibitem[{{La Barbera} {et~al}\mbox{.}(2013){La Barbera}, {Ferreras},
  {Vazdekis}, {de la Rosa}, {de Carvalho}, {Trevisan}, {Falc{\'o}n-Barroso}, \&
  {Ricciardelli}}]{LaBarbera+13_SPIDERVIII_IMF}
{La Barbera} F., {Ferreras} I., {Vazdekis} A., {de la Rosa} I.~G.,
{de
  Carvalho} R.~R., {Trevisan} M., {Falc{\'o}n-Barroso} J., {Ricciardelli} E.,
  2013, \mnras, 433, 3017

\bibitem[{{Li} {et~al}\mbox{.}(2017){Li}, {Ge}, {Mao}, {Cappellari}, {Long},
  {Li}, {Emsellem}, {Dutton}, {Li}, {Bundy}, {Thomas}, {Drory}, \&
  {Lopes}}]{Li+17_IMF}
{Li} H. {et~al.}, 2017, \apj, 838, 77

\bibitem[{{Lyubenova} {et~al}\mbox{.}(2016){Lyubenova},
  {Mart{\'{\i}}n-Navarro}, {van de Ven}, {Falc{\'o}n-Barroso}, {Galbany},
  {Gallazzi}, {Garc{\'{\i}}a-Benito}, {Gonz{\'a}lez Delgado}, {Husemann}, {La
  Barbera}, {Marino}, {Mast}, {Mendez-Abreu}, {Peletier},
  {S{\'a}nchez-Bl{\'a}zquez}, {S{\'a}nchez}, {Trager}, {van den Bosch},
  {Vazdekis}, {Walcher}, {Zhu}, {Zibetti}, {Ziegler}, {Bland-Hawthorn}, \&
  {CALIFA Collaboration}}]{Lyubenova+16}
{Lyubenova} M. {et~al.}, 2016, \mnras, 463, 3220

\bibitem[{{Macci{\`o}} {et~al}\mbox{.}(2008){Macci{\`o}}, {Dutton}, \& {van den
  Bosch}}]{Maccio+08}
{Macci{\`o}} A.~V., {Dutton} A.~A., {van den Bosch} F.~C., 2008,
\mnras, 391,
  1940

\bibitem[{{Mandelbaum} {et~al}\mbox{.}(2006){Mandelbaum}, {Seljak},
  {Kauffmann}, {Hirata}, \& {Brinkmann}}]{Mandelbaum+06}
{Mandelbaum} R., {Seljak} U., {Kauffmann} G., {Hirata} C.~M.,
{Brinkmann} J.,
  2006, \mnras, 368, 715

\bibitem[{{Maraston} {et~al}\mbox{.}(2013){Maraston}, {Pforr}, {Henriques},
  {Thomas}, {Wake}, {Brownstein}, {Capozzi}, {Tinker}, {Bundy}, {Skibba},
  {Beifiori}, {Nichol}, {Edmondson}, {Schneider}, {Chen}, {Masters}, {Steele},
  {Bolton}, {York}, {Weaver}, {Higgs}, {Bizyaev}, {Brewington}, {Malanushenko},
  {Malanushenko}, {Snedden}, {Oravetz}, {Pan}, {Shelden}, \&
  {Simmons}}]{Maraston+13_BOSS}
{Maraston} C. {et~al.}, 2013, \mnras, 435, 2764

\bibitem[{{Maraston} \& {Str{\"o}mb{\"a}ck}(2011)}]{Maraston_Stromback11}
{Maraston} C., {Str{\"o}mb{\"a}ck} G., 2011, \mnras, 418, 2785

\bibitem[{{Maraston} {et~al}\mbox{.}(2009){Maraston}, {Str{\"o}mb{\"a}ck},
  {Thomas}, {Wake}, \& {Nichol}}]{Maraston+09_LRG}
{Maraston} C., {Str{\"o}mb{\"a}ck} G., {Thomas} D., {Wake} D.~A.,
{Nichol}
  R.~C., 2009, \mnras, 394, L107

\bibitem[{{Marinoni} \& {Hudson}(2002)}]{MH02}
{Marinoni} C., {Hudson} M.~J., 2002, \apj, 569, 101

\bibitem[{{Mart\'in-Navarro} {et~al}\mbox{.}(2015){Mart\'in-Navarro},
  {Barbera}, {Vazdekis}, {Falc{\'o}n-Barroso}, \&
  {Ferreras}}]{Martin-Navarro+15_IMF_variation}
{Mart\'in-Navarro} I., {Barbera} F.~L., {Vazdekis} A.,
{Falc{\'o}n-Barroso} J.,
  {Ferreras} I., 2015, \mnras, 447, 1033

\bibitem[{{McDermid} {et~al}\mbox{.}(2014){McDermid}, {Cappellari}, {Alatalo},
  {Bayet}, {Blitz}, {Bois}, {Bournaud}, {Bureau}, {Crocker}, {Davies}, {Davis},
  {de Zeeuw}, {Duc}, {Emsellem}, {Khochfar}, {Krajnovi{\'c}}, {Kuntschner},
  {Morganti}, {Naab}, {Oosterloo}, {Sarzi}, {Scott}, {Serra}, {Weijmans}, \&
  {Young}}]{McDermid+14_IMF}
{McDermid} R.~M. {et~al.}, 2014, \apjl, 792, L37

\bibitem[{{Mehlert} {et~al}\mbox{.}(2000){Mehlert}, {Saglia}, {Bender}, \&
  {Wegner}}]{Mehlert+00}
{Mehlert} D., {Saglia} R.~P., {Bender} R., {Wegner} G., 2000,
\aaps, 141, 449

\bibitem[{{Moster} {et~al}\mbox{.}(2010){Moster}, {Somerville}, {Maulbetsch},
  {van den Bosch}, {Macci{\`o}}, {Naab}, \& {Oser}}]{Moster+10}
{Moster} B.~P., {Somerville} R.~S., {Maulbetsch} C., {van den
Bosch} F.~C.,
  {Macci{\`o}} A.~V., {Naab} T., {Oser} L., 2010, \apj, 710, 903

\bibitem[{{Napolitano} {et~al}\mbox{.}(2005){Napolitano}, {Capaccioli},
  {Romanowsky}, {Douglas}, {Merrifield}, {Kuijken}, {Arnaboldi}, {Gerhard}, \&
  {Freeman}}]{Napolitano+05}
{Napolitano} N.~R. {et~al.}, 2005, \mnras, 357, 691

\bibitem[{{Napolitano} {et~al}\mbox{.}(2010){Napolitano}, {Romanowsky}, \&
  {Tortora}}]{NRT10}
{Napolitano} N.~R., {Romanowsky} A.~J., {Tortora} C., 2010,
\mnras, 405, 2351

\bibitem[{{Navarro} {et~al}\mbox{.}(1996){Navarro}, {Frenk}, \&
  {White}}]{NFW96}
{Navarro} J.~F., {Frenk} C.~S., {White} S.~D.~M., 1996, \apj, 462,
563

\bibitem[{{Nigoche-Netro} {et~al}\mbox{.}(2016){Nigoche-Netro}, {Ramos-Larios},
  {Lagos}, {Ruelas-Mayorga}, {de la Fuente}, {Kemp}, {Navarro}, {Corral}, \&
  {Hidalgo-G{\'a}mez}}]{Nigoche-Netro+16}
{Nigoche-Netro} A. {et~al.}, 2016, \mnras, 462, 951

\bibitem[{{Oguri} {et~al}\mbox{.}(2014){Oguri}, {Rusu}, \& {Falco}}]{Oguri+14}
{Oguri} M., {Rusu} C.~E., {Falco} E.~E., 2014, \mnras, 439, 2494

\bibitem[{{Padmanabhan} {et~al}\mbox{.}(2004){Padmanabhan}, {Seljak},
  {Strauss}, {Blanton}, {Kauffmann}, {Schlegel}, {Tremonti}, {Bahcall},
  {Bernardi}, {Brinkmann}, {Fukugita}, \& {Ivezi{\'c}}}]{Padmanabhan+04}
{Padmanabhan} N. {et~al.}, 2004, \na, 9, 329

\bibitem[{{Poci} {et~al}\mbox{.}(2017){Poci}, {Cappellari}, \&
  {McDermid}}]{Poci+17_slope}
{Poci} A., {Cappellari} M., {McDermid} R.~M., 2017, \mnras, 467,
1397

\bibitem[{{Posti} {et~al}\mbox{.}(2014){Posti}, {Nipoti}, {Stiavelli}, \&
  {Ciotti}}]{Posti+14}
{Posti} L., {Nipoti} C., {Stiavelli} M., {Ciotti} L., 2014,
\mnras, 440, 610

\bibitem[{{Remus} {et~al}\mbox{.}(2013){Remus}, {Burkert}, {Dolag},
  {Johansson}, {Naab}, {Oser}, \& {Thomas}}]{Remus+13}
{Remus} R.-S., {Burkert} A., {Dolag} K., {Johansson} P.~H., {Naab}
T., {Oser}
  L., {Thomas} J., 2013, \apj, 766, 71

\bibitem[{{Remus} {et~al}\mbox{.}(2017){Remus}, {Dolag}, {Naab}, {Burkert},
  {Hirschmann}, {Hoffmann}, \& {Johansson}}]{Remus+17}
{Remus} R.-S., {Dolag} K., {Naab} T., {Burkert} A., {Hirschmann}
M., {Hoffmann}
  T.~L., {Johansson} P.~H., 2017, \mnras, 464, 3742

\bibitem[{{Roche} {et~al}\mbox{.}(2010){Roche}, {Bernardi}, \&
  {Hyde}}]{Roche+10}
{Roche} N., {Bernardi} M., {Hyde} J., 2010, \mnras, 407, 1231

\bibitem[{{Ruff} {et~al}\mbox{.}(2011){Ruff}, {Gavazzi}, {Marshall}, {Treu},
  {Auger}, \& {Brault}}]{Ruff+11}
{Ruff} A.~J., {Gavazzi} R., {Marshall} P.~J., {Treu} T., {Auger}
M.~W.,
  {Brault} F., 2011, \apj, 727, 96

\bibitem[{{Rusin} {et~al}\mbox{.}(2003){Rusin}, {Kochanek}, \&
  {Keeton}}]{Rusin+03}
{Rusin} D., {Kochanek} C.~S., {Keeton} C.~R., 2003, \apj, 595, 29

\bibitem[{{Ruszkowski} \& {Springel}(2009)}]{RS09}
{Ruszkowski} M., {Springel} V., 2009, \apj, 696, 1094

\bibitem[{{Saglia} {et~al}\mbox{.}(2000){Saglia}, {Maraston}, {Greggio},
  {Bender}, \& {Ziegler}}]{Saglia+00}
{Saglia} R.~P., {Maraston} C., {Greggio} L., {Bender} R.,
{Ziegler} B., 2000,
  \aap, 360, 911

\bibitem[{{Saglia} {et~al}\mbox{.}(2010){Saglia}, {S{\'a}nchez-Bl{\'a}zquez},
  {Bender}, {Simard}, {Desai}, {Arag{\'o}n-Salamanca}, {Milvang-Jensen},
  {Halliday}, {Jablonka}, {Noll}, {Poggianti}, {Clowe}, {De Lucia},
  {Pell{\'o}}, {Rudnick}, {Valentinuzzi}, {White}, \& {Zaritsky}}]{Saglia+10}
{Saglia} R.~P. {et~al.}, 2010, \aap, 524, A6

\bibitem[{{Salpeter}(1955)}]{Salpeter55}
{Salpeter} E.~E., 1955, \apj, 121, 161

\bibitem[{{Sarzi} {et~al}\mbox{.}(2006){Sarzi}, {Falc{\'o}n-Barroso}, {Davies},
  {Bacon}, {Bureau}, {Cappellari}, {de Zeeuw}, {Emsellem}, {Fathi},
  {Krajnovi{\'c}}, {Kuntschner}, {McDermid}, \& {Peletier}}]{Sarzi+06_SAURONV}
{Sarzi} M. {et~al.}, 2006, \mnras, 366, 1151

\bibitem[{{Schlafly} \& {Finkbeiner}(2011)}]{Schlafly_Finkbeiner11}
{Schlafly} E.~F., {Finkbeiner} D.~P., 2011, \apj, 737, 103

\bibitem[{{Shankar} \& {Bernardi}(2009)}]{Shankar_Bernardi09}
{Shankar} F., {Bernardi} M., 2009, \mnras, 396, L76

\bibitem[{{Shankar} {et~al}\mbox{.}(2010){Shankar}, {Marulli}, {Bernardi},
  {Dai}, {Hyde}, \& {Sheth}}]{Shankar+10}
{Shankar} F., {Marulli} F., {Bernardi} M., {Dai} X., {Hyde} J.~B.,
{Sheth}
  R.~K., 2010, \mnras, 403, 117

\bibitem[{{Shetty} \& {Cappellari}(2014)}]{Shetty_Cappellari14}
{Shetty} S., {Cappellari} M., 2014, \apjl, 786, L10

\bibitem[{{Shu} {et~al}\mbox{.}(2015){Shu}, {Bolton}, {Brownstein},
  {Montero-Dorta}, {Koopmans}, {Treu}, {Gavazzi}, {Auger}, {Czoske},
  {Marshall}, \& {Moustakas}}]{Shu+15_SLACSXII}
{Shu} Y. {et~al.}, 2015, \apj, 803, 71

\bibitem[{{Smith} {et~al}\mbox{.}(2015){Smith}, {Lucey}, \&
  {Conroy}}]{Smith+15_SINFONI}
{Smith} R.~J., {Lucey} J.~R., {Conroy} C., 2015, \mnras, 449, 3441

\bibitem[{{Sonnenfeld} {et~al}\mbox{.}(2017){Sonnenfeld}, {Nipoti}, \&
  {Treu}}]{Sonnenfeld+17_IMF}
{Sonnenfeld} A., {Nipoti} C., {Treu} T., 2017, \mnras, 465, 2397

\bibitem[{{Sonnenfeld} {et~al}\mbox{.}(2013){Sonnenfeld}, {Treu}, {Gavazzi},
  {Suyu}, {Marshall}, {Auger}, \& {Nipoti}}]{Sonnenfeld+13_SL2S_IV}
{Sonnenfeld} A., {Treu} T., {Gavazzi} R., {Suyu} S.~H., {Marshall}
P.~J.,
  {Auger} M.~W., {Nipoti} C., 2013, \apj, 777, 98

\bibitem[{{Sonnenfeld} {et~al}\mbox{.}(2015){Sonnenfeld}, {Treu}, {Marshall},
  {Suyu}, {Gavazzi}, {Auger}, \& {Nipoti}}]{Sonnenfeld+15_SL2SV}
{Sonnenfeld} A., {Treu} T., {Marshall} P.~J., {Suyu} S.~H.,
{Gavazzi} R.,
  {Auger} M.~W., {Nipoti} C., 2015, \apj, 800, 94

\bibitem[{{Sparks} \& {Jorgensen}(1993)}]{Sparks_Jorgensen93}
{Sparks} W.~B., {Jorgensen} I., 1993, \aj, 105, 1753

\bibitem[{{Spiniello} {et~al}\mbox{.}(2015){Spiniello}, {Barnab{\`e}},
  {Koopmans}, \& {Trager}}]{Spiniello+15_IMF_vs_density}
{Spiniello} C., {Barnab{\`e}} M., {Koopmans} L.~V.~E., {Trager}
S.~C., 2015,
  \mnras, 452, L21

\bibitem[{{Spiniello} {et~al}\mbox{.}(2012){Spiniello}, {Trager}, {Koopmans},
  \& {Chen}}]{Spiniello+12}
{Spiniello} C., {Trager} S.~C., {Koopmans} L.~V.~E., {Chen} Y.~P.,
2012, \apjl,
  753, L32

\bibitem[{{Swindle} {et~al}\mbox{.}(2011){Swindle}, {Gal}, {La Barbera}, \& {de
  Carvalho}}]{SPIDER-V}
{Swindle} R., {Gal} R.~R., {La Barbera} F., {de Carvalho} R.~R.,
2011, \aj,
  142, 118

\bibitem[{{Thomas} {et~al}\mbox{.}(2005){Thomas}, {Maraston}, {Bender}, \&
  {Mendes de Oliveira}}]{Thomas+05}
{Thomas} D., {Maraston} C., {Bender} R., {Mendes de Oliveira} C.,
2005, \apj,
  621, 673

\bibitem[{{Thomas} {et~al}\mbox{.}(2013){Thomas}, {Steele}, {Maraston},
  {Johansson}, {Beifiori}, {Pforr}, {Str{\"o}mb{\"a}ck}, {Tremonti}, {Wake},
  {Bizyaev}, {Bolton}, {Brewington}, {Brownstein}, {Comparat}, {Kneib},
  {Malanushenko}, {Malanushenko}, {Oravetz}, {Pan}, {Parejko}, {Schneider},
  {Shelden}, {Simmons}, {Snedden}, {Tanaka}, {Weaver}, \&
  {Yan}}]{Thomas+13_BOSS}
{Thomas} D. {et~al.}, 2013, \mnras, 431, 1383

\bibitem[{{Thomas} {et~al}\mbox{.}(2007){Thomas}, {Saglia}, {Bender}, {Thomas},
  {Gebhardt}, {Magorrian}, {Corsini}, \& {Wegner}}]{ThomasJ+07}
{Thomas} J., {Saglia} R.~P., {Bender} R., {Thomas} D., {Gebhardt}
K.,
  {Magorrian} J., {Corsini} E.~M., {Wegner} G., 2007, \mnras, 382, 657

\bibitem[{{Thomas} {et~al}\mbox{.}(2009){Thomas}, {Saglia}, {Bender}, {Thomas},
  {Gebhardt}, {Magorrian}, {Corsini}, \& {Wegner}}]{ThomasJ+09}
{Thomas} J., {Saglia} R.~P., {Bender} R., {Thomas} D., {Gebhardt}
K.,
  {Magorrian} J., {Corsini} E.~M., {Wegner} G., 2009, \apj, 691, 770

\bibitem[{{Thomas} {et~al}\mbox{.}(2011){Thomas}, {Saglia}, {Bender}, {Thomas},
  {Gebhardt}, {Magorrian}, {Corsini}, {Wegner}, \& {Seitz}}]{ThomasJ+11}
{Thomas} J. {et~al.}, 2011, \mnras, 415, 545

\bibitem[{{Tortora} {et~al}\mbox{.}(2017){Tortora}, {Koopmans}, \&
  {Napolitano}}]{Tortora+17_Verlinde}
{Tortora} C., {Koopmans} L.~V.~E., {Napolitano} N.~R., 2017, ArXiv
e-prints

\bibitem[{{Tortora} {et~al}\mbox{.}(2016){Tortora}, {La Barbera}, \&
  {Napolitano}}]{TLBN16_IMF_dwarfs}
{Tortora} C., {La Barbera} F., {Napolitano} N.~R., 2016, \mnras,
455, 308

\bibitem[{{Tortora} {et~al}\mbox{.}(2012){Tortora}, {La Barbera}, {Napolitano},
  {de Carvalho}, \& {Romanowsky}}]{SPIDER-VI}
{Tortora} C., {La Barbera} F., {Napolitano} N.~R., {de Carvalho}
R.~R.,
  {Romanowsky} A.~J., 2012, \mnras, 425, 577

\bibitem[{{Tortora} {et~al}\mbox{.}(2014{\natexlab{a}}){Tortora}, {La Barbera},
  {Napolitano}, {Romanowsky}, {Ferreras}, \& {de
  Carvalho}}]{Tortora+14_DMslope}
{Tortora} C., {La Barbera} F., {Napolitano} N.~R., {Romanowsky}
A.~J.,
  {Ferreras} I., {de Carvalho} R.~R., 2014{\natexlab{a}}, \mnras, 445, 115

\bibitem[{{Tortora} {et~al}\mbox{.}(2010{\natexlab{a}}){Tortora}, {Napolitano},
  {Cardone}, {Capaccioli}, {Jetzer}, \& {Molinaro}}]{Tortora+10CG}
{Tortora} C., {Napolitano} N.~R., {Cardone} V.~F., {Capaccioli}
M., {Jetzer}
  P., {Molinaro} R., 2010{\natexlab{a}}, \mnras, 407, 144

\bibitem[{{Tortora} {et~al}\mbox{.}(2009){Tortora}, {Napolitano}, {Romanowsky},
  {Capaccioli}, \& {Covone}}]{Tortora+09}
{Tortora} C., {Napolitano} N.~R., {Romanowsky} A.~J., {Capaccioli}
M., {Covone}
  G., 2009, \mnras, 396, 1132

\bibitem[{{Tortora} {et~al}\mbox{.}(2010{\natexlab{b}}){Tortora}, {Napolitano},
  {Romanowsky}, \& {Jetzer}}]{Tortora+10lensing}
{Tortora} C., {Napolitano} N.~R., {Romanowsky} A.~J., {Jetzer} P.,
  2010{\natexlab{b}}, \apjl, 721, L1

\bibitem[{{Tortora} {et~al}\mbox{.}(2011){Tortora}, {Napolitano}, {Romanowsky},
  {Jetzer}, {Cardone}, \& {Capaccioli}}]{Tortora+11MtoLgrad}
{Tortora} C., {Napolitano} N.~R., {Romanowsky} A.~J., {Jetzer} P.,
{Cardone}
  V.~F., {Capaccioli} M., 2011, \mnras, 418, 1557

\bibitem[{{Tortora} {et~al}\mbox{.}(2014{\natexlab{b}}){Tortora}, {Napolitano},
  {Saglia}, {Romanowsky}, {Covone}, \& {Capaccioli}}]{Tortora+14_DMevol}
{Tortora} C., {Napolitano} N.~R., {Saglia} R.~P., {Romanowsky}
A.~J., {Covone}
  G., {Capaccioli} M., 2014{\natexlab{b}}, \mnras, 445, 162

\bibitem[{{Tortora} {et~al}\mbox{.}(2014{\natexlab{c}}){Tortora}, {Romanowsky},
  {Cardone}, {Napolitano}, \& {Jetzer}}]{Tortora+14_MOND}
{Tortora} C., {Romanowsky} A.~J., {Cardone} V.~F., {Napolitano}
N.~R., {Jetzer}
  P., 2014{\natexlab{c}}, \mnras, 438, L46

\bibitem[{{Tortora} {et~al}\mbox{.}(2013){Tortora}, {Romanowsky}, \&
  {Napolitano}}]{TRN13_SPIDER_IMF}
{Tortora} C., {Romanowsky} A.~J., {Napolitano} N.~R., 2013, \apj,
765, 8

\bibitem[{{Treu} {et~al}\mbox{.}(2010){Treu}, {Auger}, {Koopmans}, {Gavazzi},
  {Marshall}, \& {Bolton}}]{Treu+10}
{Treu} T., {Auger} M.~W., {Koopmans} L.~V.~E., {Gavazzi} R.,
{Marshall} P.~J.,
  {Bolton} A.~S., 2010, \apj, 709, 1195

\bibitem[{{Treu} \& {Koopmans}(2004)}]{TK04}
{Treu} T., {Koopmans} L.~V.~E., 2004, \apj, 611, 739

\bibitem[{{Trujillo} {et~al}\mbox{.}(2004){Trujillo}, {Burkert}, \&
  {Bell}}]{TBB04}
{Trujillo} I., {Burkert} A., {Bell} E.~F., 2004, \apjl, 600, L39

\bibitem[{{Trujillo} {et~al}\mbox{.}(2007){Trujillo}, {Conselice}, {Bundy},
  {Cooper}, {Eisenhardt}, \& {Ellis}}]{Trujillo+07}
{Trujillo} I., {Conselice} C.~J., {Bundy} K., {Cooper} M.~C.,
{Eisenhardt} P.,
  {Ellis} R.~S., 2007, \mnras, 382, 109

\bibitem[{{Trujillo} {et~al}\mbox{.}(2011){Trujillo}, {Ferreras}, \& {de La
  Rosa}}]{Trujillo+11}
{Trujillo} I., {Ferreras} I., {de La Rosa} I.~G., 2011, \mnras,
415, 3903

\bibitem[{{Trujillo} {et~al}\mbox{.}(2006){Trujillo}, {F{\"o}rster Schreiber},
  {Rudnick}, {Barden}, {Franx}, {Rix}, {Caldwell}, {McIntosh}, {Toft},
  {H{\"a}ussler}, {Zirm}, {van Dokkum}, {Labb{\'e}}, \& {}}]{Trujillo+06}
{Trujillo} I. {et~al.}, 2006, \apj, 650, 18

\bibitem[{{Valentinuzzi} {et~al}\mbox{.}(2010{\natexlab{a}}){Valentinuzzi},
  {Fritz}, {Poggianti}, {Cava}, {Bettoni}, {Fasano}, {D'Onofrio}, {Couch},
  {Dressler}, {Moles}, {Moretti}, {Omizzolo}, {Kj{\ae}rgaard}, {Vanzella}, \&
  {Varela}}]{Valentinuzzi+10_WINGS}
{Valentinuzzi} T. {et~al.}, 2010{\natexlab{a}}, \apj, 712, 226

\bibitem[{{Valentinuzzi} {et~al}\mbox{.}(2010{\natexlab{b}}){Valentinuzzi},
  {Poggianti}, {Saglia}, {Arag{\'o}n-Salamanca}, {Simard},
  {S{\'a}nchez-Bl{\'a}zquez}, {D'onofrio}, {Cava}, {Couch}, {Fritz}, {Moretti},
  \& {Vulcani}}]{Valentinuzzi+10_EDisCS}
{Valentinuzzi} T. {et~al.}, 2010{\natexlab{b}}, \apjl, 721, L19

\bibitem[{{van den Bosch} {et~al}\mbox{.}(2007){van den Bosch}, {Yang}, {Mo},
  {Weinmann}, {Macci{\`o}}, {More}, {Cacciato}, {Skibba}, \& {Kang}}]{vdB+07}
{van den Bosch} F.~C. {et~al.}, 2007, \mnras, 376, 841

\bibitem[{{van der Wel} {et~al}\mbox{.}(2008){van der Wel}, {Holden}, {Zirm},
  {Franx}, {Rettura}, {Illingworth}, \& {Ford}}]{vanderWel+08}
{van der Wel} A., {Holden} B.~P., {Zirm} A.~W., {Franx} M.,
{Rettura} A.,
  {Illingworth} G.~D., {Ford} H.~C., 2008, \apj, 688, 48

\bibitem[{{van Dokkum} \& {Conroy}(2010)}]{vDC10}
{van Dokkum} P.~G., {Conroy} C., 2010, \nat, 468, 940

\bibitem[{{van Dokkum} \& {Franx}(2001)}]{vanDokkum_Franx01}
{van Dokkum} P.~G., {Franx} M., 2001, \apj, 553, 90

\bibitem[{{van Dokkum} {et~al}\mbox{.}(2010){van Dokkum}, {Whitaker},
  {Brammer}, {Franx}, {Kriek}, {Labb{\'e}}, {Marchesini}, {Quadri}, {Bezanson},
  {Illingworth}, {Muzzin}, {Rudnick}, {Tal}, \& {Wake}}]{vanDokkum+10}
{van Dokkum} P.~G. {et~al.}, 2010, \apj, 709, 1018

\bibitem[{{Vulcani} {et~al}\mbox{.}(2014){Vulcani}, {Bamford},
  {H{\"a}u{\ss}ler}, {Vika}, {Rojas}, {Agius}, {Baldry}, {Bauer}, {Brown},
  {Driver}, {Graham}, {Kelvin}, {Liske}, {Loveday}, {Popescu}, {Robotham}, \&
  {Tuffs}}]{Vulcani+14}
{Vulcani} B. {et~al.}, 2014, \mnras, 441, 1340

\bibitem[{{Wegner} {et~al}\mbox{.}(2012){Wegner}, {Corsini}, {Thomas},
  {Saglia}, {Bender}, \& {Pu}}]{Wegner+12}
{Wegner} G.~A., {Corsini} E.~M., {Thomas} J., {Saglia} R.~P.,
{Bender} R., {Pu}
  S.~B., 2012, \aj, 144, 78

\bibitem[{{Weidner} {et~al}\mbox{.}(2013){Weidner}, {Ferreras}, {Vazdekis}, \&
  {La Barbera}}]{Weidner+13_giant_ell}
{Weidner} C., {Ferreras} I., {Vazdekis} A., {La Barbera} F., 2013,
\mnras, 435,
  2274

\bibitem[{{Wu} {et~al}\mbox{.}(2014){Wu}, {Gerhard}, {Naab}, {Oser},
  {Martinez-Valpuesta}, {Hilz}, {Churazov}, \& {Lyskova}}]{Wu+14}
{Wu} X., {Gerhard} O., {Naab} T., {Oser} L., {Martinez-Valpuesta}
I., {Hilz}
  M., {Churazov} E., {Lyskova} N., 2014, \mnras, 438, 2701

\bibitem[{{Xu} {et~al}\mbox{.}(2017){Xu}, {Springel}, {Sluse}, {Schneider},
  {Sonnenfeld}, {Nelson}, {Vogelsberger}, \& {Hernquist}}]{Xu+17_Illustris}
{Xu} D., {Springel} V., {Sluse} D., {Schneider} P., {Sonnenfeld}
A., {Nelson}
  D., {Vogelsberger} M., {Hernquist} L., 2017, \mnras, 469, 1824

\end{thebibliography}

\end{document}